\documentclass[11pt]{article}
\pdfoutput=1
\usepackage{jheppub,amsmath,amssymb,amsthm,mathtools,bm}
\newcommand{\Tr}{\rm Tr}
\newcommand{\dd}{\mathrm d}
\newcommand{\cR}{\mathcal R}
\newcommand{\cP}{\mathcal P}
\newcommand{\C}{\mathbb C}
\newcommand{\R}{\mathbb R}

\begin{document}
\title{Secondary invariants and non-perturbative states}
\author[a,b]{Robert de Mello Koch,}

\author[b]{ and Jo\~ao P. Rodrigues,}
\affiliation[a]{School of Science, Huzhou Normal University, Huzhou 313000, China}
\emailAdd{robert@zjhu.edu.cn}
\affiliation[b]{National Institute for Theoretical and Computational Sciences, Mandelstam Institute for Theoretical Physics, School of Physics, University of the Witwatersrand, Private Bag 3, Wits 2050, South Africa}
\emailAdd{Joao.Rodrigues@wits.ac.za}
\date{April 2026}
\abstract{At finite $N$ the ring of gauge invariant operators is not freely generated. For problems of interest in physics, these rings are Cohen--Macaulay and admit a Hironaka decomposition, in which the full invariant ring is a free module over a polynomial ring generated by the primary invariants. The module basis is given by finitely many secondary invariants. This motivates a physical picture in which the primary invariants are regarded as perturbative degrees of freedom while the secondary invariants are associated with distinguished non-perturbative states or sectors. The purpose of this study is to show that a concrete algebraic version of this picture is visible in simple zero-dimensional matrix integrals.}
\maketitle


\section{Introduction}

Understanding the structure of the algebra of gauge-invariant operators at finite $N$ is a fundamental problem in matrix models and gauge theory, since it provides a window into the structure of the Hilbert space of the theory~\cite{deMelloKoch:2025ngs,deMelloKoch:2025rkw,deMelloKoch:2026dfo}. At infinite $N$ one is guided by a simple Fock-space picture: one constructs multi-trace operators from an unrestricted set of elementary building blocks, and the resulting operator algebra is freely generated. This is how the supergravity Fock space of the gravity dual~\cite{Maldacena:1997re,Gubser:1998bc,Witten:1998qj} is recovered. At finite $N$, however, trace relations become important~\cite{Pr}. These relations remove the naive redundancy of the multi-trace description and force the invariant algebra to develop a more intricate structure.

A natural language for describing this finite-$N$ structure is provided by invariant theory\cite{Pr,PI,Sturmfels}. For the invariant rings of interest in physics, the gauge group is a linearly reductive group and the ring is proved to be Cohen--Macaulay~\cite{hrtheorem}. It therefore admits a Hironaka decomposition. Concretely, if $\cR$ denotes the invariant ring, one may choose a polynomial subring
\begin{equation}
\cP=\C[p_1,\dots,p_{N_P}],\label{eq:introP}
\end{equation}
generated by a homogeneous system of parameters\footnote{The homogeneous system of parameters is also called the set of primary invariants.}, together with a finite set of secondary invariants $s_\alpha$, $\alpha=0,\dots,N_S-1$, such that
\begin{equation}
\cR=\bigoplus_{\alpha=0}^{N_S-1} \cP\,s_\alpha.\label{eq:introHironaka}
\end{equation}
Thus the full invariant ring is a free module over the ring of primary invariants, with the module basis given by the secondary invariants. The primary and secondary invariants are polynomials in single and multi-traces of the matrix variables. Each term summed in the direct sum above stands for a set of terms
\begin{equation}
\cP\,s_\alpha=\{s_\alpha, p_1s_\alpha, p_2s_\alpha,p_1^2 s_\alpha, p_1p_2s_\alpha,\cdots\}
\end{equation}
To get some insight into the physical meaning of this structure, an especially suggestive realization of \eqref{eq:introHironaka} is provided by the $d$-matrix harmonic oscillator. Let $A_i^\dagger$, $i=1,\dots,d$, denote bosonic creation operators transforming in the adjoint of the gauge group
\begin{equation}
A_i^\dagger\longrightarrow U A_i^\dagger U^\dagger,\qquad U\in U(N).\label{eq:oscadjoint}
\end{equation}
Gauge-invariant states are obtained by acting on the Fock vacuum with invariant polynomials in the $A_i^\dagger$. If $\cR$ denotes the corresponding invariant ring, then the singlet Hilbert space may be written as
\begin{equation}
{\cal H}_{\rm singlet}=\bigl\{f(A_1^\dagger,\dots,A_d^\dagger)|0\rangle : f\in \cR\bigr\}.\label{eq:singletspace}
\end{equation}
Thus every invariant polynomial determines, after acting on the vacuum, a gauge-invariant state. Since the invariant ring admits the Hironaka decomposition \eqref{eq:introHironaka}, acting on the vacuum immediately induces a corresponding decomposition of the singlet Hilbert space
\begin{equation}
{\cal H}_{\rm singlet}=\bigoplus_{\alpha=0}^{N_S-1} \cP(A^\dagger)\,s_\alpha(A^\dagger)|0\rangle.\label{eq:singletHironaka}
\end{equation}
In this form the primaries and secondaries acquire a direct physical interpretation~~\cite{deMelloKoch:2025ngs,deMelloKoch:2025rkw}. The secondary invariants generate a finite set of distinguished ``seed'' states
\begin{equation}
|s_\alpha\rangle=s_\alpha(A^\dagger)|0\rangle,
\label{eq:seedstates}
\end{equation}
while the primary invariants act freely on each such seed state to generate a tower of states. In other words, the Hironaka decomposition organizes the singlet Hilbert space itself as a finite direct sum of sectors built by acting with the primaries on a set of secondary states.

This structure is strongly suggestive. It is natural to think of the secondary states as distinguished non-perturbative states, or backgrounds, and of the primaries as generating a Fock-like space of excitations built on each such non-perturbative state. We do not assume this interpretation as an established fact but rather, we take it as physical motivation. Many of the matrix models that we study are expected to have holographic dual descriptions that admit black hole solutions. The corresponding black hole microstates, which at least for models dual to asymptotically AdS backgrounds should dominate the Hilbert space, appear as non-perturbative states. The entropy of a black hole is given by $S_{BH}=A_h/(4G_N)$ where the Newton constant $G_N$ is given by $G_N=1/N^2$ and the horizon area $A_h$ is a fixed number independent of $N$. To explain this entropy, the number of microstates must grow as $\sim e^{c N^2}$ with $c$ an order 1 number. This matches the growth in the number of secondary invariants~\cite{deMelloKoch:2025qeq}, lending non-trivial support that the secondary invariants are related to non-perturbative states. A main goal of this paper is to provide further evidence for this interpretation, in the simplest setting of zero dimensional matrix integrals.

Familiar lessons from semiclassical physics~\cite{Rajaraman:1975ez} suggest how this question can be explored. In many quantum systems, non-perturbative physics is organized around distinguished sectors. In a canonical description one expands around a background configuration and constructs perturbative fluctuations on top of it. In a path-integral description the same idea is reflected in the decomposition into contributions associated with different stationary points of the action. At a schematic level, if $\{\phi_\beta\}$ denotes a set of relevant stationary points, then one expects a semiclassical expansion of the form\footnote{If the saddle has a non-trivial moduli space one would need to integrate the collective coordinates over this moduli space, with an appropriate measure.}
\begin{equation}
Z\sim \sum_\beta \,e^{-S[\phi_\beta]/\hbar}\,\int d\eta\,\, e^{-S_{\rm pert}[\eta]},
\label{eq:semiclassicalschematic}
\end{equation}
where the remaining path integral is for the perturbative fluctuations $\eta$ propagating in that sector and $S_{\rm pert}$ is obtained by expanding the action about the $\beta^{\rm th}$ saddle. The path integral of the model, written in terms of the original matrix degrees of freedom, has a loop expansion parameter given by the couplings of the model. After the path integral is expressed in terms of invariant variables, it features $1/N^2$ as the loop expansion parameter~\cite{JS1,JS2}. It is in this description that we expect to make contact with the dual gravitational physics i.e. in this description we expect the black hole microstates to appear as saddles of the path integral. The point of the present paper is to search for this structure in the simplest setting of invariant matrix integrals. Concretely, we study zero-dimensional multi-matrix integrals of the form
\begin{equation}
Z=\int \dd M_1\cdots \dd M_d\,f(M_1,\dots,M_d),\label{eq:zerodintegral}
\end{equation}
where the integrand is invariant under simultaneous conjugation,
\begin{equation}
M_a\longrightarrow U M_a U^\dagger,\qquad U\in U(N).\label{eq:simconj}
\end{equation}
We restrict throughout to $N=2$, and study the cases $d=2,3,4$, for which the relevant Hironaka decompositions are known and the change of variables can be carried out explicitly. In these examples the number of primary invariants is $1+(d-1)N^2$, and the number of secondaries depends on $d$. In particular, the two-matrix example has only one trivial secondary invariant, the three-matrix example has one trivial and one non-trivial secondary, and the four-matrix example exhibits one trivial and seven non-trivial secondaries. Our analysis proceeds by rewriting the matrix variables in terms of trace data and residual orbit data. For $N=2$ Hermitian matrices this is especially transparent because each matrix can be written in Pauli-vector form, and simultaneous conjugation acts as a common $SO(3)$ rotation on the associated vectors. The invariant quotient is therefore controlled by Gram-type data\footnote{The Gram matrix of a set of vectors is the matrix of inner products. When the inner product is real, the Gram matrix is positive semidefinite, and every positive semidefinite matrix is the Gramian matrix for some set of vectors~\cite{Gram}. See also~\cite{Rodrigues:1992ru}.}. In the two-matrix case this leads to a single branch over the primary variables. In the three-matrix case one finds, in addition to the primary Gram data, a discrete orientation sign. In the four-matrix case the even quotient is encoded by a quartic equation for an auxiliary variable $\Delta_{(4)}$, while odd cubic invariants distinguish a further double cover. Over the complexified quotient this naturally organizes the generic fiber into eight algebraic branches above primary space. In the real Hermitian problem, however, positivity constraints select a distinguished real slice of this algebraic geometry, so that not every algebraic branch need appear as a separate real branch of the original contour.

This geometric picture has a direct consequence for the matrix integral. Once the variables are changed from matrix entries to invariant data, the integral reorganizes into a finite sum of branchwise contributions. On the generic locus the matrix integral takes the schematic form
\begin{equation}
Z\sim \sum_{\beta} \int \prod_{A=1}^{N_P}\dd p_A\,f_\beta(p),
\label{eq:branchschematic}
\end{equation}
where the $p_A$ are the primary invariants and the label $\beta$ runs over the sheets of the finite algebraic cover. Here $N_P$ denotes the number of primary invariants. In the real Hermitian setting only those branches compatible with the positivity constraints are realized directly on the original contour. Nevertheless, the algebraic cover remains physically relevant: the primaries furnish the continuous variables integrated over on each branch, while the secondaries encode the discrete fiber data that distinguishes the different algebraic sectors lying above the same point in primary space. In this sense the Hironaka decomposition becomes physically interesting. It does not merely provide a convenient basis for the invariant ring. It implies that, over primary space, the quotient is generically a finite algebraic cover, and that the secondary invariants encode the discrete data of that cover.

In every matrix-model example considered here, the number of branches of the algebraic cover is equal to the number of secondary invariants. It is natural to conjecture that this is a general feature. To test this idea in a setting where the analysis remains tractable for arbitrary $N$, we turn to a much simpler system:  $N$ bosons moving in two dimensions. Here the permutation group $S_N$, acting by exchange of particles, is treated as a gauge symmetry. The invariant ring of the model has $2N$ primary invariants and $N!$ secondary invariants \cite{deMelloKoch:2025eqt,Domokos}. For this system one can verify explicitly that the integral over the full configuration space is rewritten as an integral over the primary invariants with an accompanying sum over $N!$ branches. This gives substantial evidence that, more generally, the number of branches of the algebraic cover is controlled by the number of secondary invariants.

This already suggests a natural non-perturbative interpretation. Perturbation theory is intrinsically local in configuration space, and therefore probes only one local branch of the invariant geometry at a time. The full invariant ring, by contrast, captures the global algebraic structure of the quotient, including the finite fiber over fixed primary data. The secondary invariants should therefore not be thought of as extra perturbative oscillators. Rather, they distinguish the inequivalent algebraic branches lying above the same point in primary space. In the matrix integral this leads naturally to a sum over branches: the primaries are the continuous variables governing fluctuations within each branch, while the secondaries encode the discrete branch data needed to reconstruct the full cover. This is the precise sense in which they naturally signal non-perturbative structure.

The branch structure appears naturally once the theory is expressed in invariant variables. This has an obvious resemblance to the semiclassical decomposition of a path integral, although one should keep two distinct questions apart. The first concerns the exact organization of the quotient: even before any approximation is made, the invariant ring and the exact change of variables already reveal a finite set of algebraic branches. The second concerns dynamics: whether these branches are realized as physically relevant saddles of the path integral. In the examples we study, the answer takes a suggestive form: the same branch data are reproduced as the critical locus of a natural action in invariant space. This does not by itself establish that each algebraic branch contributes as an independent real saddle of the original Hermitian integral\footnote{The relevance of saddles that do not lie on the original integration contour is by no means a new idea. The classic example is that of instantons, which are saddles of the Euclidean action rather than of the original action, yet nevertheless encode physically important non-perturbative effects.}. What it does establish is that the branch structure is not merely formal. It is encoded in a genuine variational problem, and for that reason it gives concrete support to the interpretation in which the primary invariants describe perturbative fluctuations, while the secondary invariants carry discrete sector data of a genuinely non-perturbative character.

The paper is organized as follows. In sections~\ref{sec:two}, \ref{sec:three} and \ref{sec:four} we explicitly work out the changes of variables for two, three and four $2\times 2$ Hermitian matrices, respectively, and derive the corresponding reduced measures and domains in invariant variables. In section~\ref{SNexample} we perform the same analysis for a system of bosons in two dimensions. In section~\ref{sec:hironaka} we explain why the Hironaka decomposition naturally leads to a finite algebraic cover over primary space, and why the secondaries should be viewed as coordinates on the discrete fiber rather than as additional continuous variables. In section~\ref{sec:saddles} we explain how the algebraic branch structure can be realized as the critical locus of an invariant-space action, and how this supports the interpretation of the secondaries as labels of non-perturbative sectors. In section~\ref{sec:discussion} we discuss the broader physical meaning of this picture, its limitations, and natural directions for further work. The appendices collect a number of technical results, including checks of the invariant system used in the main text and the details of a relevant integral computation.

\section{Change of variables for two $2\times 2$ Hermitian matrices}\label{sec:two}

Our first example is a two-matrix model. We begin with a matrix integral over two Hermitian matrices $M_1$ and $M_2$, and rewrite it in terms of invariant variables. Since the number of invariants is smaller than the number of matrix elements, this change of variables necessarily introduces, in addition to the invariants themselves, a set of redundant variables parametrizing the symmetry directions that are being quotiented out. By expanding the matrices in the basis of Pauli matrices, the geometry underlying the invariant description becomes completely explicit. This allows us to rewrite the original matrix integral as an integral directly over the invariant variables. We then test the resulting expression by showing that it reproduces the correct correlation functions.
 
Our starting point is the integral
\begin{equation}
Z=\int \dd M_1\,\dd M_2\,f(M_1,M_2),
\label{eq:twoI}
\end{equation}
where the integrand $f(M_1,M_2)$ is invariant under $M_i\to U M_i U^\dagger$. This model only has the trivial secondary invariant $s_0=1$. It has five primary invariants, which may be taken to be~\cite{FHL,deMelloKoch:2025ngs}
\begin{equation}
p_1=\Tr M_1,\qquad p_2=\Tr M_2,\qquad p_3=\Tr M_1^2,\qquad p_4=\Tr(M_1M_2),\qquad p_5=\Tr M_2^2.\label{eq:twoPrimariesList}
\end{equation}
Each matrix has four real parameters, so the pair $(M_1,M_2)$ has eight real parameters. The Pauli matrices, together with the identity matrix provide a basis for $2\times2$ Hermitian matrices. Thus we can expand each matrix as
\begin{equation}
M_i=\frac{p_i}{2}\,{\bf 1}+\vec m_i\cdot \vec\sigma,\qquad i=1,2.\label{eq:twoPauliForm}
\end{equation}
Writing
\begin{equation}
M_1=\begin{pmatrix} a_1 & \xi_1+i\eta_1 \\ \xi_1-i\eta_1 & b_1 \end{pmatrix},\qquad
M_2=\begin{pmatrix} a_2 & \xi_2+i\eta_2 \\ \xi_2-i\eta_2 & b_2 \end{pmatrix}.
\label{eq:twomatricescomponents}
\end{equation}
we have $p_1=a_1+b_1$, $p_2=a_2+b_2$ and
\begin{equation}
\vec m_1=\left(\xi_1,-\eta_1,\frac{a_1-b_1}{2}\right),\qquad 
\vec m_2=\left(\xi_2,-\eta_2,\frac{a_2-b_2}{2}\right).\label{eq:twoVectors}
\end{equation}
Thus the eight real variables in $M_1$ and $M_2$ can be traded for $(p_1,p_2,\vec m_1,\vec m_2)\in \R^2\times \R^3\times \R^3$. This linear change of variables has unit Jacobian
\begin{equation}
\dd a_1\,\dd b_1\,\dd\xi_1\,\dd\eta_1\,\dd a_2\,\dd b_2\,\dd\xi_2\,\dd\eta_2
=\dd p_1\,\dd p_2\,\dd^3m_1\,\dd^3m_2.
\label{eq:twoJacLinear}
\end{equation}
Using the identities $\Tr(\sigma_i)=0$, $\Tr(\sigma_i\sigma_j)=2\delta_{ij}$ and
\begin{equation}
(\vec a\cdot\vec\sigma)(\vec b\cdot\vec\sigma)=(\vec a\cdot\vec b)\,{\bf 1}+i(\vec a\times\vec b)\cdot\vec\sigma,
\label{eq:PauliIdentity}
\end{equation}
we easily find
\begin{eqnarray}
p_1&=&\Tr M_1,\qquad p_2\,\,=\,\,\Tr M_2,\qquad p_3\,\,=\,\,\Tr M_1^2=\frac{p_1^2}{2}+2|\vec m_1|^2,\cr\cr
p_5&=&\Tr M_2^2=\frac{p_2^2}{2}+2|\vec m_2|^2,\qquad p_4\,\,=\,\,\Tr(M_1M_2)=\frac{p_1p_2}{2}+2\vec m_1\cdot\vec m_2.
\end{eqnarray}
Thus the five primary invariants determine
\begin{equation}
|\vec m_1|^2=\frac{2p_3-p_1^2}{4},\qquad |\vec m_2|^2=\frac{2p_5-p_2^2}{4},\qquad
\vec m_1\cdot\vec m_2=\frac{2p_4-p_1p_2}{4}. \label{eq:twogramentries}
\end{equation}
This gives a simple geometric description of the complete data contained in the pair $(M_1,M_2)$ that is invariant under simultaneous conjugation: under simultaneous conjugation the trace parts $p_1,p_2$ are unchanged and the vectors $\vec m_1,\vec m_2$ undergo a common $SO(3)$ rotation,
\begin{equation}
\vec m_i\longrightarrow R(U)\vec m_i,\qquad R(U)\in SO(3).\label{eq:tworotation}
\end{equation}
Hence the orbit data consist of the two traces, the two lengths, and the relative angle. These are precisely the quantities encoded by the five primary invariants $p_1,\cdots ,p_5$. The remaining three real variables in $(M_1,M_2)$, which are gauge redundant, are the common orientation of the pair of vectors $(\vec{m}_1,\vec{m}_2)$ in $\R^3$. We can use this geometrical interpretation to simplify the computation of the measure for the primary invariants. First, introduce the natural variables
\begin{equation}
r_1=|\vec m_1|,\qquad r_2=|\vec m_2|,\qquad c=\cos\theta=\frac{\vec m_1\cdot\vec m_2}{r_1r_2}.\label{eq:twor1r2c}
\end{equation}
In terms of these variables the primary invariants can be written as
\begin{equation}
p_3=\frac{p_1^2}{2}+2r_1^2,\qquad p_5=\frac{p_2^2}{2}+2r_2^2, \qquad p_4=\frac{p_1p_2}{2}+2r_1r_2c.
\label{eq:twop3p4p5r}
\end{equation}
For invariant integrands, the angular decomposition is, as usual, given by
\begin{equation}
\dd^3m_1\,\dd^3m_2=r_1^2\dd r_1\,\dd\Omega_1\;r_2^2\dd r_2\,\dd\Omega_2.\label{eq:twoangdecomp}
\end{equation}
Since only the relative angle matters, the redundant angular integrals give a constant
\begin{equation}
\int \dd\Omega_1\,\dd\Omega_2=(4\pi)(2\pi)\int_{-1}^1 \dd c\label{eq:twoangconstant}
\end{equation}
so that the measure becomes
\begin{equation}
\dd^3m_1\,\dd^3m_2=8\pi^2 r_1^2r_2^2\dd r_1\,\dd r_2\,\dd c\label{eq:twoMeasureRRc}
\end{equation}
for gauge invariant functions. Now define
\begin{equation}
\alpha=p_3-\frac{p_1^2}{2}=2r_1^2,\qquad\beta=p_5-\frac{p_2^2}{2}=2r_2^2,\qquad
\gamma=p_4-\frac{p_1p_2}{2}=2r_1r_2c.\label{eq:twoalphabetagamma}
\end{equation}
It immediately follows that
\begin{eqnarray}
\dd\alpha&=&4r_1\dd r_1,\qquad\dd\beta\,\,=\,\,4r_2\dd r_2,\qquad\dd\gamma\,\,=\,\,2r_1r_2\dd c,\cr\cr
&&\qquad\Rightarrow\quad r_1^2r_2^2\dd r_1\,\dd r_2\,\dd c=\frac{1}{32}\dd\alpha\,\dd\beta\,\dd\gamma.
\label{eq:twoJacAlphaBetaGamma}
\end{eqnarray}
Since the change of variables from $(p_1,p_2,\alpha,\beta,\gamma)$ to $(p_1,p_2,p_3,p_4,p_5)$ is triangular with unit Jacobian, we simply obtain
\begin{equation}
\dd M_1\,\dd M_2=\frac{\pi^2}{4}\,\dd p_1\,\dd p_2\,\dd p_3\,\dd p_4\,\dd p_5\label{eq:twoReducedMeasure}
\end{equation}
after the redundant angular variables are integrated out. 

All that remains is to determine the integration domain in the space of primary invariants. The variables $p_1,\dots,p_5$ are not independent over all of $\R^5$. They must satisfy the positivity conditions $r_1^2\ge 0$, $r_2^2\ge 0$, which give
\begin{equation}
2p_3-p_1^2\ge 0,\qquad 2p_5-p_2^2\ge 0.\label{eq:twoPositivity}
\end{equation}
Also $|c|\le 1$, so $\gamma^2\le \alpha\beta$. In terms of traces,
\begin{equation}
(2p_4-p_1p_2)^2\le (2p_3-p_1^2)(2p_5-p_2^2).\label{eq:twoGramineq}
\end{equation}
Thus the integration region is given by
\begin{equation}
\mathcal D_2=\Bigl\{(p_1,\dots,p_5)\in\R^5:2p_3-p_1^2\ge 0,\ \ 2p_5-p_2^2\ge 0,
\ \ (2p_4-p_1p_2)^2\le (2p_3-p_1^2)(2p_5-p_2^2)\Bigr\}.\label{eq:twoDomain}
\end{equation}
This is exactly the condition that the $2\times2$ Gram matrix of the two vectors $\vec m_1,\vec m_2$ be positive semidefinite. Finally we obtain
\begin{equation}
Z=\int \dd M_1\,\dd M_2\,f(M_1,M_2)=\frac{\pi^2}{4}\,\int_{\mathcal D_2}\dd p_1\,\dd p_2\,\dd p_3\,\dd p_4\,\dd p_5\,f(p_1,p_2,p_3,p_4,p_5)\,.
\end{equation}

\subsection{Reproducing Gaussian Correlators}

To verify the validity of the change to invariant variables, we compute correlators and test that the correct values are obtained. We do this for the free theory where the computations are simplest. Note that the measure induced by the change of variables and the integration domain are independent of the potential in the original matrix integral.

Consider the integral written in terms of the original matrix variables
\begin{equation}
I_0=\int \dd M_1\,\int \dd M_2\,\,\,\mathcal N\,\,e^{-\frac12\Tr(M_1^2)-\frac12\Tr(M_2^2)}
\label{eq:twoGaussianNorm}
\end{equation}
with $\mathcal N$ fixed so that $I_0=1$. After a simple Gaussian integration we find
\begin{equation}
\mathcal N=\frac{1}{4\pi^4}.
\label{eq:twoNormValue}
\end{equation}
Using standard techniques it is easy to verify that
\begin{eqnarray}
\langle\Tr(M_1^2)^n\Tr(M_2^2)^m\rangle&=&\int \dd M_1\,\int \dd M_2\,\,\,\mathcal N\,\,e^{-\frac12\Tr(M_1^2)-\frac12\Tr(M_2^2)}\Tr(M_1^2)^n\Tr(M_2^2)^m\cr\cr
&=&2^{n+m} (n+1)!(m+1)!\label{corrorigvars}
\end{eqnarray}
Using the reduced variables, the normalization is fixed by requiring
\begin{equation}
1=\mathcal N\frac{\pi^2}{4}\int_{\mathcal D_2} \dd p_1\cdots \dd p_5\,e^{-\frac12 p_3-\frac12 p_5}.
\label{eq:twoNormalizationPrimary}
\end{equation}
Writing out the limits of integration defined by ${\mathcal D_2}$ explicitly, this becomes
\begin{align}
1&=\mathcal N\frac{\pi^2}{4}\,\int_{-\infty}^{\infty}\dd p_1\,\int_{-\infty}^{\infty}\dd p_2\,\int_{p_1^2/2}^{\infty}\dd p_3\,\int_{p_2^2/2}^{\infty}\dd p_5\,\int_{l_-}^{l_+}\dd p_4
\,e^{-\frac12 p_3-\frac12 p_5}\,.\label{eq:twoNormalizationExpanded}
\end{align}
where
\begin{equation}
l_{\pm}=\frac{p_1p_2\pm\sqrt{(2p_3-p_1^2)(2p_5-p_2^2)}}{2}\,.
\end{equation}
A simple computation fixes $\mathcal N$ to the value already obtained in \eqref{eq:twoNormValue}. The correlator check is equally simple. After performing the integral over $p_4$, which is trivial, we obtain
\begin{eqnarray}
\langle\Tr(M_1^2)^n\Tr(M_2^2)^m\rangle&=&\left(\frac{1}{4\pi}\right)^2\left[\int_{-\infty}^{\infty} dp_1\int_{p_1^2/2}^\infty dp_3 \,\,e^{-\frac{p_3}{2}}\,p_3^n\sqrt{2p_3-p_1^2}\right]\cr\cr\cr
&&\qquad\qquad\times \left[\int_{-\infty}^{\infty} dp_2\int_{p_2^2/2}^\infty dp_5 \,\,e^{-\frac{p_5}{2}}\,p_5^m\sqrt{2p_5-p_2^2}\right]\, .
\end{eqnarray}
Now using the easily verified identity
\begin{equation}
\frac{1}{4\pi}\,\int_{-\infty}^{\infty} dp\int_{p^2/2}^\infty dq \,\,e^{-\frac{q}{2}}\,q^k\sqrt{2q-p^2}\,\,=\,\,2^k (k+1)!
\end{equation}
we can perform the remaining integrals to obtain an answer that is in complete agreement with the direct computation in matrix variables given in \eqref{corrorigvars}. This is a convincing test of the invariant measure and integration domain.

\section{Change of variables for three $2\times2$ Hermitian matrices}\label{sec:three}

In this section we consider a three-matrix model. The two matrix model of the previous section had only a trivial secondary invariant, $s_0=1$, which is always present. The three matrix model is the first example with a non-trivial secondary invariant. The change of variables has an additional structure: the integral over the original matrix elements becomes an integral over a two sheeted cover of the space parametrised by the primary invariants. The non-trivial secondary invariant is sensitive to the two sheets. Apart from these new features, the analysis proceeds in complete parallel to the last section.

We again start from a matrix integral
\begin{equation}
Z=\int \dd M_1\,\dd M_2\,\dd M_3\,f(M_1,M_2,M_3)\label{eq:threeI}
\end{equation}
where the integrand $f(M_1,M_2,M_3)$ is invariant under the simultaneous conjugation $M_i\to UM_iU^\dagger$. In the above integral there are a total of twelve integration variables. This model has nine primary invariants, which we take to be~\cite{Teranishi,deMelloKoch:2025ngs}
\begin{align}
&p_1=\Tr M_1,  &p_2&=\Tr M_2,  &p_3&=\Tr M_3,\nonumber\\
&p_4=\Tr M_1^2,  &p_5&=\Tr(M_1M_2),  &p_6&=\Tr(M_1M_3),\nonumber\\
&p_7=\Tr M_2^2,  &p_8&=\Tr M_3^2, &p_9&=\Tr(M_2M_3),\label{eq:threePrimariesList}
\end{align}
and a single non-trivial secondary invariant of degree three. We take the two secondaries to be~\cite{Teranishi,deMelloKoch:2025ngs}
\begin{equation}
s_0=1\qquad\qquad s_1=\Tr(M_1M_2M_3)\,.\label{twomatsecondaries}
\end{equation}
In the real Hermitian problem the genuinely new orbit-separating data is the sign of the imaginary part of this cubic trace. As for the two matrix model, we expand
\begin{equation}
M_i=\frac{p_i}{2}\,{\bf 1}+\vec m_i\cdot\vec\sigma,
\qquad i=1,2,3,
\label{eq:threePauliForm}
\end{equation}
replacing the twelve integration parameters in $M_1,M_2,M_3$ by the twelve parameters in $p_1,p_2,p_3,\vec m_1,\vec m_2,\vec m_3$. This linear change of variables has unit Jacobian so that
\begin{equation}
\dd M_1\,\dd M_2\,\dd M_3=\dd p_1\,\dd p_2\,\dd p_3\,\dd^3m_1\,\dd^3m_2\,\dd^3m_3. \label{eq:threeLinearMeasure}
\end{equation}
Next, using the formula $\Tr(M_iM_j)=\frac{p_ip_j}{2}+2\vec m_i\cdot\vec m_j$, we obtain
\begin{align}
p_4&=\Tr M_1^2=\frac{p_1^2}{2}+2|\vec m_1|^2,
& p_5&=\Tr(M_1M_2)=\frac{p_1p_2}{2}+2\vec m_1\cdot\vec m_2,
\nonumber\\
p_6&=\Tr(M_1M_3)=\frac{p_1p_3}{2}+2\vec m_1\cdot\vec m_3,
& p_7&=\Tr M_2^2=\frac{p_2^2}{2}+2|\vec m_2|^2,
\nonumber\\
p_8&=\Tr M_3^2=\frac{p_3^2}{2}+2|\vec m_3|^2,
& p_9&=\Tr(M_2M_3)=\frac{p_2p_3}{2}+2\vec m_2\cdot\vec m_3.
\label{eq:threeTracesExpanded}
\end{align}
Thus the six quantities $p_4,\dots,p_9$ determine the $3\times 3$ Gram matrix
\begin{equation}
G=
\begin{pmatrix}
|\vec{m}_1|^2 & &\vec{m}_1\cdot\vec{m}_2 & &\vec{m}_1\cdot\vec{m_3} \\
\vec{m}_1\cdot\vec{m}_2 & & |\vec{m}_2|^2 & &\vec{m}_2\cdot\vec{m_3} \\
\vec{m}_1\cdot\vec{m_3}& & \vec{m}_2\cdot\vec{m_3}& & |\vec{m}_3|^2
\end{pmatrix}
=\frac14
\begin{pmatrix}
2p_4-p_1^2 & 2p_5-p_1p_2 & 2p_6-p_1p_3 \\
2p_5-p_1p_2 & 2p_7-p_2^2 & 2p_9-p_2p_3 \\
2p_6-p_1p_3 & 2p_9-p_2p_3 & 2p_8-p_3^2
\end{pmatrix}.
\label{eq:threeGram}
\end{equation}
Our nine primary variables are exactly the three trace parts and the six entries of the Gram matrix of $(\vec m_1,\vec m_2,\vec m_3)$. Just as was the case for two matrices, this geometric interpretation will prove useful. For two vectors in $\R^3$, the Gram data determines the pair up to an $SO(3)$ rotation. For three vectors this is no longer true: the Gram matrix determines the triple only up to an $O(3)$ transformation, not up to $SO(3)$. The extra invariant is the oriented volume
\begin{equation}
\tau=\vec m_1\cdot(\vec m_2\times \vec m_3).\label{eq:tauDef}
\end{equation}
Under a common $SO(3)$ rotation, $\tau$ is invariant but under a reflection, it changes sign. In the matrix language, we have\footnote{On the Hermitian slice we have $\Tr(M_1M_2M_3)=\frac14p_1p_2p_3+p_1\vec{m}_2\cdot\vec{m}_3+p_2\vec{m}_1\cdot\vec{m}_3+p_3\vec{m}_1\cdot\vec{m}_2+2i\tau$ so the real part of the secondary invariant is already determined by the primaries.}
\begin{equation}
{\rm Im}\Tr(M_1M_2M_3)=2\tau\label{eq:ImTraceTau}
\end{equation}
which connects directly to the non-trivial secondary invariant. Moreover,
\begin{equation}
\tau^2=\det G.\label{eq:tau2detG}
\end{equation}
Thus the magnitude of $\tau$ is already determined by $p_1,\dots,p_9$, but its sign is not. Therefore $(p_1,\dots,p_9)$ gives a generic two-to-one parametrization of the orbit space. To get a faithful orbit description one must also keep the discrete sign
\begin{equation}
\varepsilon=\operatorname{sign}(\tau)=\pm 1.\label{eq:epsDef}
\end{equation}
If the integrand depends only on the nine primary invariants, it is blind to this sign and the two branches simply contribute equally. When the integrand depends on the non-trivial secondary invariant, the contributions from the two branches are different.

The three matrices we integrate over correspond to a total of twelve integration variables. The primaries, or equivalently, $p_i$ and $\vec{m}_i$ with $i=1,2,3$ provide nine integration variables. Take the remaining three continuous variables to be Euler angles $\Omega=(\alpha,\beta,\gamma)$ for a common $SO(3)$ frame. Concretely, choose an oriented orthonormal frame $(\hat e_1,\hat e_2,\hat e_3)$ and write
\begin{equation}
\vec m_1=r_1\hat e_1,\qquad\vec m_2=a\hat e_1+b\hat e_2,\qquad\vec m_3=c\hat e_1+d\hat e_2+e\hat e_3,
\label{eq:threeAdaptedCoords}
\end{equation}
which trades the three vectors $\vec{m}_i$ for the six variables $(r_1,a,b,c,d,e)$ and the three Euler angles which encode the common rotation of the whole triple. There is now a sign ambiguity: $e\to -e$ leaves the Gram matrix unchanged and flips the orientation of the triple. With the parametrization \eqref{eq:threeAdaptedCoords}, we easily find
\begin{equation}
\dd^3m_1\,\dd^3m_2\,\dd^3m_3=r_1^2b\,\dd r_1\,\dd a\,\dd b\,\dd c\,\dd d\,\dd e\,\dd\mu_{SO(3)}\,.
\label{eq:threeVectorMeasureAdapted}
\end{equation}
The integral over the Euler angles gives
\begin{equation}
\int_{SO(3)}\dd\mu_{SO(3)}=8\pi^2.
\label{eq:threeSO3norm}
\end{equation}
as usual. The Gram entries are
\begin{align}
&g_{11}=r_1^2, &g_{12}&=r_1a, &g_{13}&=r_1c,\nonumber\\ 
&g_{22}=a^2+b^2, &g_{23}&=ac+bd, &g_{33}&=c^2+d^2+e^2.
\label{eq:threeGramEntriesAdapted}
\end{align}
A direct computation of the Jacobian gives
\begin{equation}
\left|\frac{\partial(g_{11},g_{12},g_{13},g_{22},g_{23},g_{33})}{\partial(r_1,a,b,c,d,e)}\right|=8r_1^3b^2|e|.
\label{eq:threeJacobianAdapted}
\end{equation}
Since $\det G=r_1^2b^2e^2$, this can be written as
\begin{equation}
\dd^3m_1\,\dd^3m_2\,\dd^3m_3=\sum_{\varepsilon=\pm 1}\frac{1}{8\sqrt{\det G}}\,\dd g_{11}\,\dd g_{12}\,
\dd g_{13}\,\dd g_{22}\,\dd g_{23}\,\dd g_{33}\,\dd\mu_{SO(3)}.\label{eq:threeMeasureGfull}
\end{equation}
The sum over $\varepsilon$ is precisely the sum over the two signs of $e$, or equivalently the two signs of $\tau$. Therefore
\begin{equation}
\dd M_1\,\dd M_2\,\dd M_3=\dd p_1\,\dd p_2\,\dd p_3\sum_{\varepsilon=\pm 1}\frac{1}{8\sqrt{\det G}}
\,\dd^6G\,\dd\mu_{SO(3)}.\label{eq:threeFullMeasureBeforeP}
\end{equation}
All that remains now is to convert from the Gramm entries to the primaries $p_4,\cdots,p_9$ and to carefully spell out the integration domain. The map from variables $p_1,p_2,p_3,G$ to $p_1,\dots,p_9$ is triangular, with Jacobian
\begin{equation}
\dd p_1\cdots \dd p_9=64\,\dd p_1\,\dd p_2\,\dd p_3\,\dd^6G.
\label{eq:threeJacPtoG}
\end{equation}
Define $\Delta_{(3)}=\det (4G)$. In terms of $\Delta_{(3)}$ we have
\begin{equation}
\det G=\frac{\Delta_{(3)}}{64},\qquad\qquad\sqrt{\det G}=\frac{\sqrt{\Delta_{(3)}}}{8}.\label{eq:threeDetGDelta}
\end{equation}
Thus, the total measure is the sum over the two branches
\begin{equation}
\dd M_1\,\dd M_2\,\dd M_3=\sum_{\varepsilon=\pm1}\frac{1}{64\sqrt{\Delta_{(3)}}}\,\dd p_1\cdots \dd p_9\,\dd\mu_{SO(3)}.\label{eq:threeMeasureWithSO3}
\end{equation}
After integrating over $SO(3)$ one obtains\footnote{When writing this formula we have in mind correlator computations. For example, in the free theory the integrand is $e^{-\frac12\Tr M_1^2-\frac12\Tr M_2^2-\frac12\Tr M_3^2}O(M_1,M_2,M_3)$ with $O(M_1,M_2,M_3)$ given by an invariant polynomial in $M_1,M_2,M_3$, describing the correlator of interest. In terms of the primary invariants, the exponential becomes $e^{-\frac12p_4-\frac12p_7-\frac12p_8}$. The Hironaka decomposition guarantees that $O(M_1,M_2,M_3)$ can be written as a polynomial linear in the secondary invariants with coefficients that are polynomials in the primary invariants.}
\begin{eqnarray}
Z&=&\int \dd M_1\,\dd M_2\,\dd M_3\,f(M_1,M_2,M_3)\cr\cr
&=&\sum_{\varepsilon=\pm1}
\int_{G\succeq 0} \frac{\pi^2}{8\sqrt{\Delta_{(3)}}}\,\dd p_1\cdots \dd p_9\,f(p_1,\dots,p_9,s_0,s_1).
\label{eq:threeReducedMeasureBranches}
\end{eqnarray}
Finally the integration domain is exactly the Gram-domain condition $G\succeq 0$. Concretely, this condition can be written as
\begin{align}
2p_4-p_1^2\ge 0,\qquad 2p_7-p_2^2\ge 0,\qquad 2p_8-p_3^2&\ge 0,\label{eq:threeDiagIneq}\\
(2p_4-p_1^2)(2p_7-p_2^2)-(2p_5-p_1p_2)^2&\ge 0,\label{eq:threeMinor12}\\
(2p_4-p_1^2)(2p_8-p_3^2)-(2p_6-p_1p_3)^2&\ge 0,\label{eq:threeMinor13}\\
(2p_7-p_2^2)(2p_8-p_3^2)-(2p_9-p_2p_3)^2&\ge 0,\label{eq:threeMinor23}\\
\Delta_{(3)}&\ge 0.\label{eq:threeDetIneq}
\end{align}

\subsection{Reproducing Gaussian Correlators}

As we did for the two matrix model, we compute a number of correlators to verify that the transformation to invariant variables is correct. We start from the integral written in terms of the original variables
\begin{equation}
I_0=\int \dd M_1\,\dd M_2\,\dd M_3\,\mathcal N\,e^{-\frac12\Tr M_1^2-\frac12\Tr M_2^2-\frac12\Tr M_3^2}
\label{eq:threeGaussianNorm}
\end{equation}
with $\mathcal N$ fixed so that $I_0=1$. After a simple computation we find
\begin{equation}
\mathcal N=\frac{1}{8\pi^6}.
\label{eq:threeNormValue}
\end{equation}
Again, using standard methods it is easy to verify that
\begin{eqnarray}
\langle\Tr(M_1^2)^n\Tr(M_2^2)^m\Tr(M_3^2)^p\rangle&=&\int \dd M_1\int \dd M_2\int \dd M_3\,\mathcal N\,\,e^{-\frac12\Tr(M_1^2)-\frac12\Tr(M_2^2)-\frac12\Tr(M_3^2)}\cr\cr
&&\qquad\times \Tr(M_1^2)^n\Tr(M_2^2)^m\Tr(M_3^2)^p\cr\cr
&=&2^{n+m+p} (n+1)!(m+1)!(p+1)!\label{anothercorrorigvars}
\end{eqnarray}
Finally, since an odd integral vanishes, we have $\langle\Tr(M_1 M_2 M_3)^{2p+1}\rangle=0$.

We now want to recover these results directly in terms of the invariant variables. The computation is a very close parallel to the two matrix case. The normalization condition is
\begin{equation}
1=\mathcal N\frac{\pi^2}{8}\sum_{\varepsilon=\pm}\int_{\mathcal D_3}\frac{\dd p_1\cdots \dd p_9}{\sqrt{\Delta_{(3)}}}e^{-\frac12p_4-\frac12p_7-\frac12p_8},\label{eq:threeNormPrimary}
\end{equation}
where $\mathcal D_3$ is the Gram domain. The integrand does not depend on $s_1$ so there is no dependence on the branch. Thus, the branch sum gives a factor of 2. Writing things out explicitly, we have
\begin{eqnarray}
1&=&\mathcal N\frac{\pi^2}{4}\int_{-\infty}^\infty dp_1\int_{-\infty}^\infty dp_2\int_{-\infty}^\infty dp_3 \int_{p_1^2/2}^\infty dp_4\int_{p_2^2/2}^\infty dp_7\int_{p_3^2/2}^\infty dp_8\cr\cr\cr
&&\times\int_{\frac{p_1 p_2-\sqrt{(2p_4-p_1^2)(2p_7-p_2^2)}}{2}}^{\frac{p_1 p_2+\sqrt{(2p_4-p_1^2)(2p_7-p_2^2)}}{2}}\, dp_5 \int_{\frac{p_1 p_3-\sqrt{(2p_4-p_1^2)(2p_8-p_3^2)}}{2}}^{\frac{p_1 p_3+\sqrt{(2p_4-p_1^2)(2p_8-p_3^2)}}{2}}\, dp_6\cr\cr
&&\times\int_{\frac{p_2 p_3-\sqrt{(2p_8-p_3^2)(2p_7-p_2^2)}}{2}}^{\frac{p_2 p_3+\sqrt{(2p_8-p_3^2)(2p_7-p_2^2)}}{2}}\, dp_9\,\,\frac{1}{\sqrt{\Delta_{(3)}}}\,\,e^{-\frac12p_4-\frac12p_7-\frac12p_8}\,\,\theta(\Delta_{(3)}),
\end{eqnarray}
where $\theta(\Delta_{(3)})$ enforces $\Delta_{(3)}\ge 0$ and with
\begin{eqnarray}
\Delta_{(3)}&=&(2p_4-p_1^2)(2p_7-p_2^2)(2p_8-p_3^2)+2(2p_5-p_1p_2)(2p_6-p_1p_3)(2p_9-p_2p_3)\cr\cr
&&\quad-(2p_4-p_1^2)(2p_9-p_2p_3)^2-(2p_7-p_2^2)(2p_6-p_1p_3)^2-(2p_8-p_3^2)(2p_5-p_1p_2)^2\,.
\nonumber
\end{eqnarray}
To perform the normalization integral, it is simplest to trade the primary variables $p_4,\ldots,p_9$ for entries of the Gram matrix. Define
\begin{align}
\alpha_1&=2p_4-p_1^2, &\alpha_2&=2p_7-p_2^2, &\alpha_3&=2p_8-p_3^2\nonumber\\
\alpha_4&=2p_5-p_1p_2, &\alpha_5&=2p_6-p_1p_3, &\alpha_6&=2p_9-p_2p_3.
\end{align}
The inverse transformation is
\begin{align}
p_4&=\frac{p_1^2+\alpha_1}{2}, &p_7&=\frac{p_2^2+\alpha_2}{2}, &p_8&=\frac{p_3^2+\alpha_3}{2},\nonumber\\
p_5&=\frac{p_1p_2+\alpha_4}{2}, &p_6&=\frac{p_1p_3+\alpha_5}{2}, &p_9&=\frac{p_2p_3+\alpha_6}{2},
\end{align}
and the determinant becomes
\begin{equation}
\Delta_{(3)}=\alpha_1\alpha_2\alpha_3+2\alpha_4\alpha_5\alpha_6-\alpha_1 \alpha_6^2-\alpha_2 \alpha_5^2-\alpha_3 \alpha_4^2.\label{eq:Delta3alphabetachixyz}
\end{equation}
The Jacobian for this linear change of variables is
\begin{equation}
\dd p_4\,\dd p_5\,\dd p_6\,\dd p_7\,\dd p_8\,\dd p_9=\frac{1}{64}\,\dd^6\alpha\label{eq:threeJacShifted}
\end{equation}
Thus \eqref{eq:threeNormPrimary} becomes
\begin{align}
1&=\mathcal N\frac{\pi^2}{4}\,\frac{1}{64}\int_{-\infty}^{\infty}\dd p_1\int_{-\infty}^{\infty}\dd p_2\int_{-\infty}^{\infty}\dd p_3\,e^{-\frac14(p_1^2+p_2^2+p_3^2)}\int_{\Gamma_3}\frac{\dd^6\alpha}
{\sqrt{\Delta_{(3)}}}\,e^{-\frac14(\alpha_1+\alpha_2+\alpha_3)}\,.\label{eq:threeNormShifted}
\end{align}
The integration domain $\Gamma_3$ is $\alpha_1,\alpha_2,\alpha_3\ge 0$, $\alpha_4^2\le \alpha_1\alpha_2$, $\alpha_5^2\le \alpha_1\alpha_3$ and $\Delta_{(3)}\ge 0$. The integrals over $p_1,p_2,p_3$ are elementary Gaussian integrals. Doing them we have
\begin{equation}
1=\mathcal N\,\frac{\pi^{7/2}}{32}\,\mathcal I_3,\qquad\mathcal I_3\equiv
\int_{\Gamma_3}\frac{\dd^6\alpha}{\sqrt{\Delta_{(3)}}}\,e^{-\frac14(\alpha_1+\alpha_2+\alpha_3)}.\label{eq:defI3}
\end{equation}
To compute $\mathcal I_3$ start by writing
\begin{equation}
\Delta_{(3)}=-\alpha_1\left(\alpha_6-\frac{\alpha_4\alpha_5}{\alpha_1}\right)^2
+\frac{(\alpha_1\alpha_2-\alpha_4^2)(\alpha_1\alpha_3-\alpha_5^2)}{\alpha_1}.
\label{eq:threeCompleteSquare}
\end{equation}
Thus, for $\alpha_1>0$, the allowed range of $\alpha_6$ is
\begin{equation}
\alpha_6^-\le \alpha_6\le \alpha_6^+,\qquad \alpha_6^\pm=\frac{\alpha_4\alpha_5\pm \sqrt{(\alpha_1\alpha_2-\alpha_4^2)(\alpha_1\alpha_3-\alpha_5^2)}}{\alpha_1}.
\label{eq:zpm}
\end{equation}
The set $\alpha_1=0$ has measure zero and does not affect the integral. The $\alpha_6$ integral is
\begin{align}
\int_{\alpha_6^-}^{\alpha_6^+}\frac{\dd \alpha_6}{\sqrt{\Delta_{(3)}}}=\frac{1}{\sqrt{\alpha_1}}
\int_{-\sqrt{\frac{(\alpha_1\alpha_2-\alpha_4^2)(\alpha_1\alpha_3-\alpha_5^2)}{\alpha_1}}}^{\sqrt{\frac{(\alpha_1\alpha_2-\alpha_4^2)(\alpha_1\alpha_3-\alpha_5^2)}{\alpha_1}}}
\frac{\dd u}{\sqrt{\frac{(\alpha_1\alpha_2-\alpha_4^2)(\alpha_1\alpha_3-\alpha_5^2)}{\alpha_1}-u^2}}
=\frac{\pi}{\sqrt{\alpha_1}},\label{eq:zIntegralPi}
\end{align}
where in the first step we set $u=\sqrt{\alpha_1}\left(\alpha_6-\frac{\alpha_4\alpha_5}{\alpha_1}\right)$. Therefore
\begin{align}
\mathcal I_3&=\pi\int_0^\infty\dd\alpha_1\int_0^\infty\dd\alpha_2\int_0^\infty\dd\alpha_3\,
e^{-\frac14(\alpha_1+\alpha_2+\alpha_3)}\int_{-\sqrt{\alpha_1\alpha_2}}^{\sqrt{\alpha_1\alpha_2}}\dd \alpha_4
\int_{-\sqrt{\alpha_1\alpha_3}}^{\sqrt{\alpha_1\alpha_3}}\dd \alpha_5\,\frac{1}{\sqrt{\alpha_1}}\notag\\
&=4\pi\int_0^\infty\dd\alpha_1\int_0^\infty\dd\alpha_2\int_0^\infty\dd\alpha_3\,
e^{-\frac14(\alpha_1+\alpha_2+\alpha_3)}\sqrt{\alpha_1\alpha_2\alpha_3}.\label{eq:I3reduced}
\end{align}
The remaining integrals are now straightforward and we find complete agreement with \eqref{eq:threeNormValue}.

The computation of correlation functions involves computing
\begin{equation}
\mathcal N\frac{\pi^2}{4}\int_{\mathcal D_3}\frac{\dd p_1\cdots \dd p_9}{\sqrt{\Delta_{(3)}}}e^{-\frac12p_4-\frac12p_7-\frac12p_8} p_4^n p_7^m p_8^p\, .
\end{equation}
These are easily evaluated using exactly the change of variables we described in complete detail above. The result is in perfect agreement with \eqref{anothercorrorigvars}. Finally, consider
\begin{equation}
\langle\Tr(M_1M_2M_3)^{2p+1}\rangle=\mathcal N\frac{\pi^2}{8}\sum_{\varepsilon=\pm}\int_{\mathcal D_3}\frac{\dd p_1\cdots \dd p_9}{\sqrt{\Delta_{(3)}}}e^{-\frac12p_4-\frac12p_7-\frac12p_8}\,s_1^{2p+1}
\end{equation}
On the Hermitian slice,
\begin{equation}
s_1=\Tr(M_1M_2M_3)=A(p)+2i\tau,
\end{equation}
where $A(p)$ is a branch-independent real function of the primaries and $\tau$ flips sign between branches. So the two branch values are $s_1=A(p)\pm 2i\sqrt{\det G}$, which are complex conjugates of one another. Therefore, the sum over the branches produces the real quantity $(A+2i\sqrt{\det G})^{2p+1}+(A-2i\sqrt{\det G})^{2p+1}$. In particular, there are no odd powers of $\sqrt{\det G}$ remaining after the branch sum. The symmetry $M_1\to -M_1$, $M_2\to -M_2$ and $M_3\to -M_3$ of the Gaussian measure then forces the integral to vanish, so that we recover $\langle\Tr(M_1M_2M_3)^{2p+1}\rangle=0$. 

\section{Change of variables for four $2\times2$ Hermitian matrices}\label{sec:four}

Our final example is a four-matrix model, for which the invariant ring contains eight secondary sectors: one trivial secondary invariant and seven non-trivial ones. In invariant variables, the matrix integral becomes an eight-sheeted cover of the space parametrized by the primary invariants. The secondary invariants once again furnish the discrete information required to distinguish these sheets, and hence to identify the branch of the integral being described.

Consider the matrix integral
\begin{equation}
Z=\int \dd M_1\,\dd M_2\,\dd M_3\,\dd M_4\,\,f(M_1,M_2,M_3,M_4)\label{eq:fourI}
\end{equation}
where the integrand $f(M_1,M_2,M_3,M_4)$ is invariant under simultaneous conjugation of all four matrices $M_i\to UM_iU^\dagger$. The integration is over a total of sixteen independent matrix elements. This model has thirteen primary invariants and eight secondary invariants. As in our previous examples, the interpretation of the invariants is most transparent in terms of the entries of the Gram matrix $G=(Q_{ij})_{1\le i,j\le 4}$, where the quadratic invariants are
\begin{equation}
Q_{ij}=\vec m_i\cdot\vec m_j=\frac14\bigl(2\Tr(M_iM_j)-\Tr(M_i)\Tr(M_j)\bigr),\qquad 1\le i\le j\le 4.\label{eq:QijDef}
\end{equation}
and $\vec{m}_i$ are defined by the expansion of $M_i$ in terms of Pauli matrices
\begin{equation}
M_i=\frac{t_i}{2}\,{\bf 1}+\vec m_i\cdot\vec\sigma,\qquad t_i=\Tr M_i,\qquad \vec m_i\in\R^3,\qquad i=1,2,3,4.
\label{eq:fourPauliForm}
\end{equation}
We will use the following thirteen primary invariants
\begin{align}
p_1&=\Tr(M_1), & p_2&=\Tr(M_2), &p_3&=\Tr(M_3), &p_4&=\Tr(M_4), \nonumber\\
p_5&=Q_{12}, & p_6&=Q_{13}, &p_7&=Q_{14}, &p_8&=Q_{23},\nonumber \\
p_9&=Q_{24}, & p_{10}&=Q_{34}, &p_{11}&=Q_{11}-Q_{22},\nonumber\\
p_{12}&=Q_{22}-Q_{33}, & p_{13}&=Q_{33}-Q_{44}. \label{eq:fourPrimariesExplicit}
\end{align}
To describe the secondary invariants, introduce
\begin{equation}
\Delta_{(4)}=Q_{11}+Q_{22}+Q_{33}+Q_{44}.\label{eq:DeltaDef}
\end{equation}
Using $p_{11},p_{12},p_{13}$ and $\Delta_{(4)}$, the four diagonal Gram entries can be written as
\begin{align}
Q_{11}&=\frac14\left(\Delta_{(4)}+3p_{11}+2p_{12}+p_{13}\right),&Q_{22}&=\frac14\left(\Delta_{(4)}-p_{11}+2p_{12}+p_{13}\right),\nonumber\\
Q_{33}&=\frac14\left(\Delta_{(4)}-p_{11}-2p_{12}+p_{13}\right), &Q_{44}&=\frac14\left(\Delta_{(4)}-p_{11}-2p_{12}-3p_{13}\right).\label{eq:Q44recon}
\end{align}
Thus the full Gram matrix $G(p,\Delta_{(4)})$ becomes the following explicit $4\times 4$ matrix
\begin{equation}
G=
\begin{pmatrix}
Q_{11} & p_5 & p_6 & p_7 \\
p_5 & Q_{22} & p_8 & p_9 \\
p_6 & p_8 & Q_{33} & p_{10} \\
p_7 & p_9 & p_{10} & Q_{44}
\end{pmatrix}.
\end{equation}
Since the vectors $\vec m_i$ lie in $\R^3$ they are linearly dependent and the Gram matrix $G\succeq 0$ obeys the constraint
\begin{equation}
\Phi(p,\Delta_{(4)})=\det G(p,\Delta_{(4)})=0.\label{eq:PhiConstraint}
\end{equation}
Since $\Delta_{(4)}$ enters only through the diagonal elements of the Gram matrix, it is clear that $\Phi(p,\Delta_{(4)})$ is a quartic polynomial in $\Delta_{(4)}$, with leading term
\begin{equation}
\Phi(p,\Delta_{(4)})=\frac{1}{256}\Delta_{(4)}^4+\cdots.\label{eq:PhiLeading}
\end{equation}
This makes it clear, as discussed below, that the even quotient over primary space is explicitly a quartic cover. To give the secondary invariants we also need the canonical cubic invariants
\begin{equation}
T_{ijk}=\vec m_i\cdot(\vec m_j\times \vec m_k)=\frac{1}{2i}\Tr(X_iX_jX_k),\qquad 1\le i<j<k\le 4,\label{eq:TijkDef}
\end{equation}
where $X_i=M_i-\frac{t_i}{2}\,{\bf 1}$. We take the secondary invariants to be
\begin{align}
s_0&=1,  &s_1&=\Delta_{(4)}, &s_2&=T_{123}, &s_3&=T_{124},\nonumber\\
s_4&=T_{134}, &s_5&=T_{234}, &s_6&=\Delta_{(4)}^2, &s_7&=\Delta_{(4)}^3.\label{eq:fourSecondaries}
\end{align}
The invariants for this model have also been constructed in~\cite{Teranishi2,Domokos2}. Our choice in  \eqref{eq:fourPrimariesExplicit} and \eqref{eq:fourSecondaries} differs from that of~\cite{Teranishi2,Domokos2},  and is more convenient for our purposes. We explain in Appendix \ref{invariantbases} how we have confirmed this system of invariants. The change of variables from the original matrix variables to $t_i$ and $\vec{m}_i$ is a linear change of variables and the flat measure factorizes as
\begin{equation}
\dd M_1\,\dd M_2\,\dd M_3\,\dd M_4
=\dd t_1\,\dd t_2\,\dd t_3\,\dd t_4\,\dd^3m_1\,\dd^3m_2\,\dd^3m_3\,\dd^3m_4.
\label{eq:fourLinearMeasure}
\end{equation}

As we have already emphasized, after the change of variables the primary invariants define a ramified algebraic covering of the quotient. To develop this further, fix a point $p=(p_1,\dots,p_{13})$ in primary space. Then the variable $\Delta_{(4)}$ is constrained by the quartic equation $\Phi(p,\Delta_{(4)})=0$. Generically there are four algebraic choices,
\begin{equation}
\Delta_{(4),1}(p),\ \Delta_{(4),2}(p),\ \Delta_{(4),3}(p),\ \Delta_{(4),4}(p).\label{eq:fourRoots}
\end{equation}
These give the four even algebraic sheets of the complexified quotient. On the real Hermitian contour one must in addition impose the positivity constraints on the Gram matrix, and these determine which of the algebraic sheets are realized directly on the original integration domain. On each sheet consider the cofactors of $G(p,\Delta_{(4),r}(p))$, given by
\begin{equation}
C_{ij}^{(r)}(p)=(-1)^{i+j}\det G^{(r)}_{[ji]},
\label{eq:cofactorsDef}
\end{equation}
where $G^{(r)}_{[ji]}$ is the $3\times 3$ minor obtained by deleting row $j$ and column $i$.
On the generic patch where $C_{44}^{(r)}(p)>0$, the cubic invariants on the two orientation branches are
\begin{align}
T_{123}^{(r,\varepsilon)}(p)&=\varepsilon\sqrt{C_{44}^{(r)}(p)},
&T_{124}^{(r,\varepsilon)}(p)&=-\varepsilon\frac{C_{34}^{(r)}(p)}{\sqrt{C_{44}^{(r)}(p)}},\nonumber\\
T_{134}^{(r,\varepsilon)}(p)&=\varepsilon\frac{C_{24}^{(r)}(p)}{\sqrt{C_{44}^{(r)}(p)}},
&T_{234}^{(r,\varepsilon)}(p)&=-\varepsilon\frac{C_{14}^{(r)}(p)}{\sqrt{C_{44}^{(r)}(p)}}.\label{cubicconstraints}
\end{align}
Once the primaries and a choice of the $\Delta_{(4)r}$ sheet are fixed, the remaining ambiguity is an overall sign. This residual $\mathbb Z_2$ freedom is the overall orientation of the four vectors $\vec{m}_i\in\mathbb{R}^3$. All of the cubic invariants are odd under orientation reversal $\vec{m}_i\to R(U)\vec{m}_i$ with $\det R=-1$, so they all flip sign together explaining why a single sign $\varepsilon$ determines all of them. This is manifest in the linear relations among the cubic invariants
\begin{eqnarray}
T_{124}&=&\det(\vec{m}_1,\vec{m}_2,\vec{m}_4)\,\,=\,\,c_3 T_{123}\cr\cr 
T_{134}&=&\det(\vec{m}_1,\vec{m}_3,\vec{m}_4)\,\,=\,\,-c_2 T_{123}\cr\cr 
T_{234}&=&\det(\vec{m}_2,\vec{m}_3,\vec{m}_4)\,\,=\,\,c_1 T_{123}.
\end{eqnarray}
So once $T_{123}$ is chosen, the other three cubic invariants are fixed by coefficients determined from the Gram matrix. There is one common ambiguous sign and this common sign can be taken to be the sign of $T_{123}$, which is how we chose the orientation sign $\varepsilon$. Thus, over a generic primary point $p$ the complexified quotient is naturally organized into eight algebraic branches,
\begin{equation}
\bigl\{(p,\Delta_r(p),\varepsilon):r=1,2,3,4,\ \varepsilon=\pm1\bigr\},\label{eq:eightbranches}
\end{equation}
with the understanding that the real Hermitian contour is obtained by intersecting this algebraic cover with the positivity domain.

We will now determine the integration measure in terms of the invariant variables. Start from the original matrix integral \eqref{eq:fourI}. On the generic patch $\vec m_1,\vec m_2,\vec m_3$ are linearly independent. Using this triple introduce adapted coordinates
\begin{equation}
\vec m_1=r_1\hat e_1,\qquad\vec m_2=a\hat e_1+b\hat e_2,\qquad
\vec m_3=c\hat e_1+d\hat e_2+\varepsilon e\hat e_3,\label{eq:fourAdapted123}
\end{equation}
with $r_1>0$, $b>0$, $e>0$, $\varepsilon=\pm1$, and $(\hat e_1,\hat e_2,\hat e_3)\in SO(3)$. A simple computation gives
\begin{equation}
\dd^3m_1\,\dd^3m_2\,\dd^3m_3=r_1^2b\,\dd r_1\,\dd a\,\dd b\,\dd c\,\dd d\,\dd e\,\dd\mu_{SO(3)}.
\label{eq:four123measure}
\end{equation}
The $3\times 3$ Gram matrix
\begin{equation}
A=
\begin{pmatrix}
Q_{11}&Q_{12}&Q_{13}\\
Q_{12}&Q_{22}&Q_{23}\\
Q_{13}&Q_{23}&Q_{33}
\end{pmatrix}
\label{eq:Adef}
\end{equation}
has elements given by
\begin{align}
Q_{11}&=r_1^2, &Q_{12}&=r_1a, &Q_{13}&=r_1c,\nonumber\\
Q_{22}&=a^2+b^2, &Q_{23}&=ac+bd, &Q_{33}&=c^2+d^2+e^2.\label{eq:Aentries}
\end{align}
A direct computation of the Jacobian gives
\begin{equation}
\left|\frac{\partial(Q_{11},Q_{12},Q_{13},Q_{22},Q_{23},Q_{33})}{\partial(r_1,a,b,c,d,e)}\right|=8r_1^3b^2e.
\label{eq:fourAJac}
\end{equation}
Also $\det A=r_1^2b^2e^2$. Therefore, for fixed $\varepsilon$ we find
\begin{equation}
\dd^3m_1\,\dd^3m_2\,\dd^3m_3=\frac{1}{8\sqrt{\det A}}\,\dd^6A\,\dd\mu_{SO(3)}.
\label{eq:four123measureA}
\end{equation}
Use the fact that $\vec m_1,\vec m_2,\vec m_3$ form a basis to write
\begin{equation}
\vec m_4=c_1\vec m_1+c_2\vec m_2+c_3\vec m_3.
\label{eq:m4expand}
\end{equation}
In terms of the pair of vectors
\begin{equation}
\nu=\begin{pmatrix} Q_{14}\\Q_{24}\\Q_{34}\end{pmatrix},\qquad
c=\begin{pmatrix} c_1\\c_2\\c_3\end{pmatrix}\label{eq:nuandc}
\end{equation}
we can write
\begin{equation}
Q_{44}=c^TAc=\nu^TA^{-1}\nu,\qquad{\rm where}\qquad \nu=Ac.\label{eq:nuAc}
\end{equation}
We then easily obtain the measure for integrating over $\vec{m}_4$ as follows
\begin{equation}
\dd^3m_4=|\det(\vec m_1,\vec m_2,\vec m_3)|\,\dd^3c=\sqrt{\det A}\,\dd^3c.
\label{eq:d3m4c}
\end{equation}
Using the fact that $\nu=Ac$, we find
\begin{equation}
\dd^3\nu=(\det A)\,\dd^3c,\qquad\dd^3c=\frac{\dd^3\nu}{\det A}\qquad\Rightarrow\qquad \dd^3m_4=\frac{\dd^3\nu}{\sqrt{\det A}}.\label{eq:d3m4nu}
\end{equation}
Multiplying by the three-vector measure gives, branch by branch,
\begin{equation}
\dd^3m_1\,\dd^3m_2\,\dd^3m_3\,\dd^3m_4=\frac{\dd^6A\,\dd^3\nu\,\dd\mu_{SO(3)}}{8\det A}.\label{eq:fourMeasureAnu}
\end{equation}
To write this in a slightly more convenient form, note the full Gram matrix can be written as
\begin{equation}
G=\begin{pmatrix}A & \nu \\ \nu^T & Q_{44}\end{pmatrix}\label{eq:blockGram}
\end{equation}
so that the Schur complement formula implies
\begin{equation}
\det G=(\det A)\left(Q_{44}-\nu^TA^{-1}\nu\right).\label{eq:SchurComplement}
\end{equation}
Since $A\succ0$ on the patch, we have
\begin{equation}
\delta(\det G)=\frac{1}{\det A}\delta\left(Q_{44}-\nu^TA^{-1}\nu\right) \label{eq:deltaSchur}
\end{equation}
which implies that
\begin{equation}
\frac{\dd^6A\,\dd^3\nu}{\det A}=\dd^6A\,\dd^3\nu\,\dd Q_{44}\,\delta(\det G).
\label{eq:measureWithDelta}
\end{equation}
Noting that $\dd^6A\,\dd^3\nu\,\dd Q_{44}=\prod_{1\le i\le j\le4}\dd Q_{ij}$ the branch-resolved vector measure becomes
\begin{equation}
\dd^3m_1\,\dd^3m_2\,\dd^3m_3\,\dd^3m_4=\frac18\left(\prod_{1\le i\le j\le 4}\dd Q_{ij}\right)\,\dd\mu_{SO(3)}\,\Theta(G\succeq0)\,\delta(\det G),\label{eq:vectorMeasureQ}
\end{equation}
The factor $\Theta(G\succeq0)$ must be included because not every symmetric matrix arises as a Gram matrix. In the end we find
\begin{equation}
Z=\pi^2\sum_{\varepsilon=\pm1}\int\left(\prod_{i=1}^4\dd t_i\right)
\left(\prod_{1\le i\le j\le 4}\dd Q_{ij}\right)\Theta(G\succeq0)\,\delta(\det G)\,F_\varepsilon(t,Q),\label{eq:IFFirstQ}
\end{equation}
where to obtain this formula we have integrated over the redundant $SO(3)$. Now change variables from the ten quadratic coordinates $Q_{ij}$ to $(p_5,\cdots,p_{13},\Delta_{(4)})$. The Jacobian of this linear change of variables is
\begin{equation}
\left|\frac{\partial(p_5,p_6,p_7,p_8,p_9,p_{10},p_{11},p_{12},p_{13},\Delta_{(4)})}{\partial(Q_{11},Q_{12},Q_{13},Q_{14},Q_{22},Q_{23},Q_{24},Q_{33},Q_{34},Q_{44})}\right|=4,\label{eq:fourLinearJac}
\end{equation}
which implies that
\begin{equation}
\dd^{10}Q=\frac14\dd^9p\,\dd\Delta_{(4)}\,.\label{eq:d10Qto9pDelta}
\end{equation}
Using the delta function identity
\begin{equation}
\delta\bigl(\Phi(p,\Delta_{(4)})\bigr)=\sum_{r=1}^4\frac{\delta\bigl(\Delta_{(4)}-\Delta_{(4)r}(p)\bigr)}{\left|\partial_{\Delta_{(4)}}\Phi\bigl(p,\Delta_{(4)r}(p)\bigr)\right|},\label{eq:deltaPhiRoots}
\end{equation}
valid on the generic locus where the roots are distinct, we obtain the explicit thirteen-primary integration formula
\begin{equation}
Z=\frac{\pi^2}{4}\int \dd^{13}p\sum_{r=1}^4\sum_{\varepsilon=\pm1}
\frac{\Theta_r(p)}{\left|\partial_{\Delta_{(4)}}\Phi\bigl(p,\Delta_{(4)r}(p)\bigr)\right|}
f_{r,\varepsilon}(p_1,\cdots,s_7),\label{eq:four13primaryformula}
\end{equation}
where $\Theta_r(p)=1$ if $G\bigl(p,\Delta_r(p)\bigr)\succeq0$ and vanishes otherwise. We use $f_{r,\varepsilon}(p_1,\cdots,s_7)$ to denote the integrand evaluated on branch $r$ with orientation $\varepsilon$. This is the explicit branchwise integral in terms of the thirteen primaries.

Finally, let us spell out the integration domain. The cleanest explicit limits for the primary integrations come from solving the principal-minor inequalities in nested form. To begin, we must fix a branch $r$ and define
\begin{equation}
Q^{(r)}_{11}=\frac14\Bigl(\Delta_{(4)r}+3p_{11}+2p_{12}+p_{13}\Bigr),\qquad
Q^{(r)}_{22}=\frac14\Bigl(\Delta_{(4)r}-p_{11}+2p_{12}+p_{13}\Bigr),
\end{equation}
\begin{equation}
Q^{(r)}_{33}=\frac14\Bigl(\Delta_{(4)r}-p_{11}-2p_{12}+p_{13}\Bigr),\qquad
Q^{(r)}_{44}=\frac14\Bigl(\Delta_{(4)r}-p_{11}-2p_{12}-3p_{13}\Bigr),
\end{equation}
On the generic patch on which $Q^{(r)}_{11}>0$ and $Q^{(r)}_{22}>0$, the condition $G\bigl(p,\Delta_r(p)\bigr)\succeq 0$ can be written as a sequence of nested bounds. A convenient iterated version of \eqref{eq:four13primaryformula} is then
\begin{align}
Z&=\frac{\pi^2}{4}\sum_{r=1}^4\sum_{\varepsilon=\pm1}\int_{-\infty}^{\infty}dp_1\int_{-\infty}^{\infty}dp_2
\int_{-\infty}^{\infty}dp_3\int_{-\infty}^{\infty}dp_4\int_{\mathcal D_r^{\rm diag}} dp_{11}\,dp_{12}\,dp_{13}\notag\\
&\hspace{1cm}\times\int_{-\sqrt{Q^{(r)}_{11}Q^{(r)}_{22}}}^{\sqrt{Q^{(r)}_{11}Q^{(r)}_{22}}} dp_5
\int_{-\sqrt{Q^{(r)}_{11}Q^{(r)}_{33}}}^{\sqrt{Q^{(r)}_{11}Q^{(r)}_{33}}} dp_6
\int_{-\sqrt{Q^{(r)}_{11}Q^{(r)}_{44}}}^{\sqrt{Q^{(r)}_{11}Q^{(r)}_{44}}} dp_7\notag\\
&\hspace{1cm}\times\int_{L^{(r,-)}_8}^{L^{(r,+)}_8} dp_8\int_{L^{(r,-)}_9}^{L^{(r,+)}_9} dp_9
\int_{\max\{L^{(r,-)}_{10,134},\,L^{(r,-)}_{10,234}\}}^{\min\{L^{(r,+)}_{10,134},\,L^{(r,+)}_{10,234}\}}dp_{10}\,\,
\frac{f_{r,\varepsilon}(p_1,\cdots,s_7)}{\bigl|\partial_\Delta\Phi\bigl(p,\Delta_r(p)\bigr)\bigr|},\label{eq:4.37iter}
\end{align}
where
\begin{equation}
\mathcal D_r^{\rm diag}=\left\{(p_{11},p_{12},p_{13})\in\mathbb R^3:\Delta_{(4)r}(p)\in\mathbb R,\;
Q^{(r)}_{11}\ge0,\;Q^{(r)}_{22}\ge0,\;Q^{(r)}_{33}\ge0,\;Q^{(r)}_{44}\ge0\right\},\nonumber
\end{equation}
and the nested limits are
\begin{equation}
L^{(r,\pm)}_8=\frac{p_5p_6\pm\sqrt{\bigl(Q^{(r)}_{11}Q^{(r)}_{22}-p_5^2\bigr)\bigl(Q^{(r)}_{11}Q^{(r)}_{33}-p_6^2\bigr)}}{Q^{(r)}_{11}},\nonumber
\end{equation}
\begin{equation}
L^{(r,\pm)}_9=\frac{p_5p_7\pm\sqrt{\bigl(Q^{(r)}_{11}Q^{(r)}_{22}-p_5^2\bigr)
\bigl(Q^{(r)}_{11}Q^{(r)}_{44}-p_7^2\bigr)}}{Q^{(r)}_{11}},\nonumber
\end{equation}
\begin{equation}
L^{(r,\pm)}_{10,134}=\frac{p_6p_7\pm\sqrt{\bigl(Q^{(r)}_{11}Q^{(r)}_{33}-p_6^2\bigr)
\bigl(Q^{(r)}_{11}Q^{(r)}_{44}-p_7^2\bigr)}}{Q^{(r)}_{11}},
\nonumber
\end{equation}
\begin{equation}
L^{(r,\pm)}_{10,234}=\frac{p_8p_9\pm\sqrt{\bigl(Q^{(r)}_{22}Q^{(r)}_{33}-p_8^2\bigr)
\bigl(Q^{(r)}_{22}Q^{(r)}_{44}-p_9^2\bigr)}}{Q^{(r)}_{22}}.
\nonumber
\end{equation}
This form follows by imposing positive semidefiniteness of the branch Gram matrix $G\bigl(p,\Delta_r(p)\bigr)$ through its principal minors in the order
\begin{equation}
Q^{(r)}_{11},Q^{(r)}_{22},Q^{(r)}_{33},Q^{(r)}_{44},\quad\det G^{(r)}_{12},\det G^{(r)}_{13},\det G^{(r)}_{14},\quad
\det G^{(r)}_{123},\det G^{(r)}_{124},\det G^{(r)}_{134},\det G^{(r)}_{234}.\nonumber
\end{equation}
The final bound on $p_{10}$ is the intersection of the intervals coming from the $(134)$ and $(234)$ principal minors.

Since $\Delta_r(p)$ is only implicitly determined by $\Phi(p,\Delta)=0$, the outer domain $\mathcal D_r^{\rm diag}$ is only defined implicitly. Nevertheless, \eqref{eq:4.37iter} is a genuine iterated-integral version of the branchwise formula. On patches where $Q^{(r)}_{11}=0$ or $Q^{(r)}_{22}=0$, one should simply permute the matrix labels and use a different positive diagonal entry as pivot.

\subsection{Reproducing Gaussian correlators}

As in our previous examples, we verify that the invariant measure and integration domain reproduce the correct Gaussian correlators. Again work in the free theory, where computations are simplest. Although the branchwise formula \eqref{eq:4.37iter} gives the integral explicitly in terms of the thirteen primaries, for actual evaluation it is more convenient to use the intermediate representation \eqref{eq:IFFirstQ} in terms of the trace variables $t_i$ and the Gram matrix entries $Q_{ij}$.

Consider the Gaussian integral in terms of the original matrix variables
\begin{equation}
I_0=\int dM_1\,dM_2\,dM_3\,dM_4\; {\cal N}\,\exp\left[-\frac12\Tr(M_1^2)-\frac12\Tr(M_2^2)-\frac12\Tr(M_3^2)-\frac12\Tr(M_4^2)\right], \label{eq:4.38}
\end{equation}
with ${\cal N}$ fixed by the condition $I_0=1$. Since the four matrices are independent, a simple computation in the original matrix variables gives ${\cal N}=\frac{1}{16\pi^8}$. Again, standard Gaussian integration implies
\begin{align}
&\left\langle \Tr(M_1^2)^{n_1}\Tr(M_2^2)^{n_2}\Tr(M_3^2)^{n_3}\Tr(M_4^2)^{n_4}\right\rangle\notag\\
&\hspace{1cm}=\int dM_1\,dM_2\,dM_3\,dM_4\;{\cal N}\,e^{-\frac12\sum_{a=1}^4\Tr(M_a^2)}
\prod_{a=1}^4 \Tr(M_a^2)^{n_a}\notag\\
&\hspace{1cm}=2^{n_1+n_2+n_3+n_4}\prod_{a=1}^4 (n_a+1)!.
\label{acorransinmvars}
\end{align}

We now recover these results from the invariant description. The Gaussian weight is
\begin{equation}
-\frac12\sum_{i=1}^4 \Tr(M_i^2)=-\frac14\sum_{i=1}^4 t_i^2-\sum_{i=1}^4Q_{ii}=-\frac14\sum_{i=1}^4 t_i^2-\Tr G.
\label{eq:4.42}
\end{equation}
The normalization condition is therefore
\begin{equation}
1={\cal N}\pi^2\sum_{\varepsilon=\pm1}\int\left(\prod_{i=1}^4dt_i\right)\left(\prod_{1\le i\le j\le 4}dQ_{ij}\right)
\Theta(G\succeq 0)\,\delta(\det G)\,\exp\left[-\frac14\sum_{i=1}^4 t_i^2-\Tr G\right].\label{invrep}
\end{equation}
Since the Gaussian integrand depends only on the primary data and is blind to the orientation label $\varepsilon$, the two orientation branches contribute equally and we have
\begin{equation}
1=2{\cal N}\pi^2\left[\int_{-\infty}^{\infty}dt\;e^{-t^2/4}\right]^4\int_{G\succeq 0}
\left(\prod_{1\le i\le j\le 4}dQ_{ij}\right)\delta(\det G)\,e^{-\Tr G}.
\end{equation}
The integrals over the trace $t_i$ are elementary, so fixing the normalization reduces to evaluating the Gram integral
\begin{equation}
J_4\equiv\int_{G\succeq 0}\left(\prod_{1\le i\le j\le 4}dQ_{ij}\right)\delta(\det G)\,e^{-\Tr G}.
\end{equation}
To evaluate $J_4$, use the block decomposition introduced in \eqref{eq:blockGram} to obtain
\begin{align}
J_4=\int_{A\succ 0} d^6A\int d^3\nu\,\frac{1}{\det A}\,\exp\left[-\Tr A-\nu^TA^{-1}\nu\right]=\pi^{3/2}
\int_{A\succ 0} d^6A\,\frac{e^{-\Tr A}}{\sqrt{\det A}}.
\end{align}
The remaining integral is (see Appendix \ref{IntegralDone})
\begin{equation}
\int_{A\succ 0} d^6A\,\frac{e^{-\Tr A}}{\sqrt{\det A}}=\frac{\pi^{5/2}}{2}\label{IntRess}
\end{equation}
so that $J_4=\frac{\pi^4}{2}$. It is now simple to obtain ${\cal N}=\frac{1}{16\pi^8}$, reproducing the normalization computed in the original matrix variables.

To study correlators, introduce the weighted generating function
\begin{equation}
\mathcal Z(\lambda_1,\lambda_2,\lambda_3,\lambda_4)=\Big\langle
\exp\!\Big[\sum_{a=1}^4 \lambda_a\,\Tr(M_a^2)\Big]\Big\rangle .
\end{equation}
Using the representation in terms of invariants parallel to \eqref{invrep}, this becomes
\begin{equation}
\mathcal Z(\lambda_1,\lambda_2,\lambda_3,\lambda_4)=2{\cal N}\pi^2\left[\prod_{a=1}^4 \int_{-\infty}^{\infty} dt_a\,\exp\!\left(-\frac{1-2\lambda_a}{4}\,t_a^2\right)\right] J_4(a_1,a_2,a_3,a_4),
\end{equation}
where $a_a=1-2\lambda_a$ and
\begin{equation}
J_4(a_1,a_2,a_3,a_4)=\int_{G\succeq 0}\left(\prod_{1\le i\le j\le 4} dQ_{ij}\right)\delta(\det G)\,
\exp\left(-\sum_{a=1}^4 a_a Q_{aa}\right).
\end{equation}
Now evaluate the weighted Gram integral using the block decomposition \eqref{eq:blockGram} to obtain
\begin{equation}
J_4(a_1,a_2,a_3,a_4)=\int_{A\succ 0} d^6A\,e^{-(a_1Q_{11}+a_2Q_{22}+a_3Q_{33})}
\int d^3\nu\,\frac{1}{\det A}\,e^{-a_4\,\nu^T A^{-1}\nu}.
\end{equation}
The \(\nu\)-integral is Gaussian and easily performed. The result is
\begin{equation}
J_4(a_1,a_2,a_3,a_4)=\pi^{3/2}a_4^{-3/2}\int_{A\succ 0} d^6A\,\frac{e^{-(a_1Q_{11}+a_2Q_{22}+a_3Q_{33})}}{\sqrt{\det A}}.
\end{equation}
Introduce the variable $B$ through $A=D^{-1/2}BD^{-1/2}$, with $D={\rm diag}(a_1,a_2,a_3)$. For symmetric $3\times 3$ matrices, we have
\begin{equation}
d^6A=(a_1a_2a_3)^{-2}d^6B,\qquad\det A=(a_1a_2a_3)^{-1}\det B,\qquad\Tr(DA)=\Tr B.
\end{equation}
so that
\begin{equation}
J_4(a_1,a_2,a_3,a_4)=\frac{\pi^{3/2}}{(a_1a_2a_3a_4)^{3/2}}\int_{B\succ 0} d^6B\,\frac{e^{-\Tr B}}{\sqrt{\det B}}=
\frac{\pi^4}{2\,(a_1a_2a_3a_4)^{3/2}},
\end{equation}
where we used the integral already computed in \eqref{IntRess}. The \(t_a\)-integrals are elementary
\begin{equation}
\int_{-\infty}^{\infty} dt_a\,\exp\left(-\frac{1-2\lambda_a}{4}t_a^2\right)=\frac{2\sqrt{\pi}}{\sqrt{1-2\lambda_a}}
\end{equation}
so that we obtain
\begin{equation}
\mathcal Z(\lambda_1,\lambda_2,\lambda_3,\lambda_4)=2{\cal N}\pi^2\prod_{a=1}^4\frac{2\sqrt{\pi}}{\sqrt{1-2\lambda_a}} \cdot\frac{\pi^4}{2\prod_{a=1}^4(1-2\lambda_a)^{3/2}}=\prod_{a=1}^4 \frac{1}{(1-2\lambda_a)^2}.
\end{equation}
Finally, differentiating at \(\lambda_a=0\) gives
\begin{equation}
\Big\langle \prod_{a=1}^4 \Tr(M_a^2)^{n_a}\Big\rangle=\left.\prod_{a=1}^4\frac{\partial^{n_a}}{\partial \lambda_a^{n_a}}\mathcal Z(\lambda_1,\lambda_2,\lambda_3,\lambda_4)\right|_{\lambda_a=0}=
2^{n_1+n_2+n_3+n_4}\prod_{a=1}^4 (n_a+1)!,
\end{equation}
which reproduces exactly the direct matrix-variable result \eqref{acorransinmvars}.

\section{An example with $S_N$ symmetry}\label{SNexample}

Next we consider a system of $N$ bosons in two dimensions, which is a theory with diagonal $S_N$ gauge symmetry acting by permutating particle labels. This model was studied in detail in~\cite{deMelloKoch:2025eqt} where results directly relevant for the present analysis were obtained. In particular, the algebra of $S_N$-invariants admits a Hironaka decomposition with $2N$ primary invariants and $N!$ secondary invariants. We will show that, upon passing from particle coordinates to invariant variables, the original integral becomes an integral over primary space with an $N!$-sheeted structure.

The configuration space is $\mathcal C_N=(\mathbb R^2)^N$, with coordinates $(x_1,y_1),\dots,(x_N,y_N)$ for the $N$ particles. The relevant gauge symmetry is the diagonal action of $S_N$, which permutes particle labels simultaneously in the two coordinate directions:
\begin{equation}
\pi\cdot \bigl((x_1,y_1),\dots,(x_N,y_N)\bigr)=
\bigl((x_{\pi^{-1}(1)},y_{\pi^{-1}(1)}),\dots,(x_{\pi^{-1}(N)},y_{\pi^{-1}(N)})\bigr).
\end{equation}
Two configurations are gauge equivalent when they are related by a simultaneous permutation of particle labels. The algebra of gauge-invariant observables is therefore the algebra of polynomial functions on $C_N$ invariant under this diagonal $S_N$ action.

The $2N$ primary invariants are given by the power sums~\cite{Domokos}
\begin{equation}
p_n^{(x)}=\sum_{i=1}^N x_i^n,\qquad p_n^{(y)}=\sum_{i=1}^N y_i^n,\qquad n=1,\dots,N.
\end{equation}
These define a map
\begin{equation}
\Phi:\mathcal C_N/S_N\longrightarrow \mathbb R^{2N},\qquad
[(x_i,y_i)]\mapsto \bigl(p_1^{(x)},\dots,p_N^{(x)},p_1^{(y)},\dots,p_N^{(y)}\bigr).
\end{equation}
Our goal is to understand the generic fiber of \(\Phi\).

The map $\Phi$ can be inverted because the first $N$ power sums $p_n^{(x)}$ determine the unordered $N$-tuple $\{x_1,\dots,x_N\}$, and similarly the $y$-power sums determine $\{y_1,\dots,y_N\}$. This follows because the power sums determine the elementary symmetric functions by Newton's identities. Explicitly, if
\begin{equation}
e_k^{(x)}=\sum_{1\le i_1<\cdots<i_k\le N}x_{i_1}\cdots x_{i_k},
\end{equation}
then the Newton identities express $e_k^{(x)}$ polynomially in the primaries $p_1^{(x)},\dots,p_k^{(x)}$. Thus the monic polynomial
\begin{equation}
P_x(t)=\prod_{i=1}^N (t-x_i)=t^N-e_1^{(x)}t^{N-1}+e_2^{(x)}t^{N-2}-\cdots+(-1)^N e_N^{(x)}
\end{equation}
is determined by the $x$-primaries. The unordered set of roots of $P_x(t)$ is the unordered $N$-tuple of $x$ coordinates. A completely parallel argument shows that the $y$-primaries determine the unordered $N$-tuple of $y$ coordinates.

To have a clean sheet count, restrict to the generic locus where both root sets are simple $\Delta_x\neq 0$, $\Delta_y\neq 0$, where the discriminants are given by
\begin{equation}
\Delta_x=\prod_{i<j}(\alpha_i-\alpha_j)^2,\qquad\Delta_y=\prod_{i<j}(\beta_i-\beta_j)^2.
\end{equation}
Equivalently, the $x_i$ are all distinct and the $y_i$ are all distinct. 

Fix a generic primary point. Let $\{\alpha_1,\dots,\alpha_N\}$ be the distinct $x$-roots and $\{\beta_1,\dots,\beta_N\}$ the distinct $y$-roots, determined by the primaries. Take any configuration in the fiber. Since its $x$-primaries agree with the chosen primary data, its $x$-coordinates must be a permutation of the $\alpha_i$. Since we quotient by diagonal $S_N$, we may use a gauge permutation to reorder the particles so that
\begin{equation}
x_i=\alpha_i,\qquad i=1,\dots,N.
\end{equation}
After this gauge fixing, the $y_i$ must still be some permutation of the $\beta_j$. Thus there exists a unique permutation $\sigma\in S_N$ such that
\begin{equation}
y_i=\beta_{\sigma(i)},\qquad i=1,\dots,N.
\end{equation}
This implies that every orbit in the fiber has a representative of the form
\begin{equation}
(\alpha_1,\beta_{\sigma(1)}),\dots,(\alpha_N,\beta_{\sigma(N)})
\end{equation}
for some $\sigma\in S_N$. Now we want to show that distinct permutations $\sigma\neq\tau$ define distinct diagonal $S_N$-orbits. Suppose
\begin{equation}
(\alpha_1,\beta_{\sigma(1)}),\dots,(\alpha_N,\beta_{\sigma(N)})\qquad {\rm and}\qquad (\alpha_1,\beta_{\tau(1)}),\dots,(\alpha_N,\beta_{\tau(N)})
\end{equation}
are in the same orbit. Then there exists a permutation $\pi\in S_N$ such that
\begin{equation}
(\alpha_{\pi(i)},\beta_{\sigma(\pi(i))})=(\alpha_i,\beta_{\tau(i)})\qquad\text{for all }i.
\end{equation}
Comparing the first coordinates gives $\alpha_{\pi(i)}=\alpha_i$ for all $i$. Since the $\alpha_i$ are pairwise distinct, this implies $\pi(i)=i$ for all $i$, so $\pi$ is the identity permutation. Then comparing second coordinates gives $\beta_{\sigma(i)}=\beta_{\tau(i)}$ for all $i$. Since the $\beta_j$ are pairwise distinct, this implies $\sigma(i)=\tau(i)$ for all $i$, so that $\sigma=\tau$. Since there are $N!$ distinct permutations, this argument shows that the generic fiber has $N!$ elements. Equivalently, over the generic primary locus, the orbit space is an $N!$-sheeted cover of the primary space.

On the nongeneric locus where $\Delta_x=0$ or $\Delta_y=0$, some roots coincide, and some permutations become indistinguishable. So the $N!$-sheeted cover branches there.

We will now explain that, exactly as for the matrix model examples we considered above, the secondary invariants provide the data that encodes the sheet label. For a generic primary point, choose an ordering of the $x$-roots and $y$-roots:
\begin{equation}
\alpha_1,\dots,\alpha_N,\qquad\qquad\qquad \beta_1,\dots,\beta_N.
\end{equation}
The configuration belonging to the sheet labeled by $\sigma\in S_N$ is given by $(\alpha_1,\beta_{\sigma(1)}),\dots,(\alpha_N,\beta_{\sigma(N)})$. The secondary invariants are mixed invariants~\cite{deMelloKoch:2025eqt}. A mixed invariant takes the form
\begin{equation}
s_{a,b}=\sum_{i=1}^N x_i^a y_i^b
\end{equation}
On the branch labeled by $\sigma$ it takes the value
\begin{equation}
s_{a,b}^{(\sigma)}=\sum_{i=1}^N \alpha_i^a \beta_{\sigma(i)}^b.
\end{equation}
which depends on the permutation $\sigma$. This proves that the secondaries are branchwise algebraic functions that encode the sheet label $\sigma$.

Thus an integral written in terms of the original variables becomes a sum over $N!$ terms, one for each branch. To illustrate this in as transparent a setting as possible, consider moments of the Gaussian integral,
\begin{equation}
Z=\int_{\mathbb R^{2N}} \prod_{i=1}^N dx_i\,dy_i\;\exp\left[-\frac12\sum_{i=1}^N (x_i^2+y_i^2)\right]f(x_1,\cdots,y_N)\,.
\end{equation}
A completely straightforward analysis gives the invariant-space formula 
\begin{equation}
Z=\sum_{\sigma\in S_N}\int_{\Gamma} d^{2N}p\;\frac{1}{N!\sqrt{\Delta_x(p)\,\Delta_y(p)}}\;\exp\left[-\frac12\bigl(p_2^{(x)}+p_2^{(y)}\bigr)\right]\,f_{\sigma}(p_1,\cdots,p_{2N},s_0,\cdots,s_{N!-1})\label{SNmodelInvInt}
\end{equation}
where $\Delta_x(p)$ and $\Delta_y(p)$ are the discriminants of the two degree-$N$ polynomials determined by the $x$- and $y$-primaries, and $\Gamma$ is the real primary domain, given by $\Gamma=\Gamma_x\times\Gamma_y$ where
\begin{eqnarray}
\Gamma_x&=&\{(p_1^{(x)},\cdots,p_N^{(x)})\in\mathbb{R}|P_x(t){\rm \,\,has\,\,only\,\,real\,\,roots}\}\cr\cr
\Gamma_y&=&\{(p_1^{(y)},\cdots,p_N^{(y)})\in\mathbb{R}|P_y(t){\rm \,\,has\,\,only\,\,real\,\,roots}\}
\end{eqnarray}
and where $P_x(t)$ and $P_y(t)$ are the monic polynomials introduced above. The change of variables to the primary invariants produces an integral over primary space alone, with no explicit sum over $S_N$. This is because the primary variables parametrize only the base of the covering map, not the full orbit space. For generic primary data, the power sums determine the unordered sets of $x$-roots and $y$-roots, but they do not determine how these two sets are paired to form the $N$ particles. The missing information, which must be supplied before the secondary invariants can be evaluated, is precisely a permutation $\sigma\in S_N$, specifying the pairing between the $x$-roots and the $y$-roots. Thus the full orbit space is, generically, an $N!$-sheeted cover of primary space. The sum over $S_N$ is therefore not produced by the Jacobian, but it must be included to give the correct value of the integral defined by the starting expression in the original coordinates. When the integrand depends only on the primaries this refinement is trivial, since all sheets contribute equally, but it becomes essential once one introduces secondary invariants, whose branchwise values distinguish the different sheets. Thus in this example we again see that the number of branches is equal to the number of secondary invariants.

\section{Hironaka decomposition and the geometry of the quotient}\label{sec:hironaka}

The examples presented above suggest a simple general picture. After choosing a set of primary invariants, the quotient is organized over the corresponding primary space by a finite algebraic fiber. The secondary invariants encode this finite fiber. In this section we explain the precise sense in which this statement follows from the Hironaka decomposition, and why it is this finite algebraic structure is naturally tied to the secondary invariants.

It is useful to keep in mind the four-matrix example, where the geometry is explicit. The thirteen primary invariants $p_A$ define the base, while the variable
\begin{equation}
\Delta_{(4)}=Q_{11}+Q_{22}+Q_{33}+Q_{44}\label{eq:DeltaAgain}
\end{equation}
is constrained by the Gram-matrix condition
\begin{equation}
\Phi(p,\Delta_{(4)})=\det G(p,\Delta_{(4)})=0.\label{eq:PhiAgain}
\end{equation}
For generic values of the primaries this is a quartic equation in $\Delta_{(4)}$. Thus, at the level of the complexified quotient, the even sector defines a four-sheeted algebraic cover of primary space. The odd cubic invariants $T_{123},T_{124},T_{134},T_{234}$ distinguish a further double cover. In this sense the complexified quotient has generic algebraic degree eight over primary space. Specifying the secondary invariants corresponds to specifying a particular sheet of the cover. In the real Hermitian problem one must in addition impose the positivity constraints on the Gram matrix, and these constraints determine which of the algebraic branches are actually realized on the original contour. The important point is that the full algebraic quotient carries eight generic sheets and that the secondaries are the invariants that distinguish them.

The correct general framework for this statement is the Hironaka decomposition of the invariant ring. The physical content of the Hironaka decomposition is that the primary and secondary invariants play sharply different roles. The primaries behave as the natural continuous gauge-invariant variables: they parametrize the base of the quotient space and are the variables over which the invariant formulation of the theory is naturally integrated. The secondaries do not generate additional continuous directions. Instead, they encode the finite residual data that remain once the primaries have been fixed. Equivalently, the full gauge-invariant configuration space is generically a finite cover of primary space, with the secondaries distinguishing the different sheets lying above the same primary data. From the point of view of the path integral, this is precisely the structure expected of a sum over branches or sectors: the primaries describe continuous fluctuations within a chosen branch, while the secondaries specify which branch one is on. In this sense, the primaries naturally resemble perturbative variables, whereas the secondaries encode discrete sector data of a more intrinsically non-perturbative kind.

This interpretation matches the explicit examples studied above. In the two-matrix example the generic fiber is trivial and there is only the trivial secondary. In the three-matrix example the quotient carries a non-trivial double cover distinguished by the orientation sign, and there are two secondaries. In the four-matrix example the natural algebraic structure is an eight-sheeted cover of the complexified primary space: four sheets associated with the quartic equation for $\Delta$, and a further doubling distinguished by the odd cubic data. In the four matrix examples there are eight secondaries. Finally, for the $S_N$ model the quotient is an $N!$ sheeted cover and there are $N!$ secondary invariants. The Hironaka decomposition therefore does much more than provide a convenient algebraic basis. It predicts that the quotient is organized, over primary space, by a finite algebraic fiber whose generic degree, counted with multiplicity, is the number of secondaries.

There is a second, useful algebraic perspective. Let $u\in\cR$ be any invariant. Multiplication by $u$ defines a $\cP$-linear map
\begin{equation}
\mu_u:\cR\to \cR,\qquad x\mapsto ux.\label{eq:multmap}
\end{equation}
Since $\cR$ is a free $\cP$-module of rank $N_S$, this map is represented by an $N_S\times N_S$ matrix with entries in $\cP$. By the Cayley--Hamilton theorem, $u$ obeys its own characteristic equation, $\chi_u(u)=0$ where $\chi_u(t)$ is a monic polynomial of degree $N_S$ with coefficients in $\cP$. Thus every invariant is algebraic over the primary ring. For simplicity, imagine we are in a particularly favourable situation in which a single invariant $u$ separates the generic fiber. Then the entire cover is described by one polynomial equation
\begin{equation}
\chi_u(t)=0.
\label{eq:singleGeneratorEqn}
\end{equation}
Since this polynomial is of degree $N_S$ the quotient is described by an $N_S$ sheeted cover of the primary space. In general one should not expect a single invariant to separate the full generic fiber. Rather, a collection of secondary invariants are needed to resolve the different sheets. This is already visible in the four-matrix example, where the generator $\Delta_{(4)}$, which obeys the quartic equation \eqref{eq:PhiConstraint}, resolves the four even algebraic sheets, while the odd cubic secondaries distinguish the further double cover associated with the orientation data. The example with $S_N$ invariance corresponds to the favourable case in which (generically) a single secondary invariant does separate the generic fiber.

This viewpoint also clarifies the structure of the invariant integral. Let $F$ be an invariant polynomial. By the Hironaka decomposition it may be expanded uniquely as
\begin{equation}
F=\sum_{\alpha=0}^{N_S-1}f_\alpha(p)s_\alpha,
\qquad f_\alpha(p)\in\cP.
\label{eq:HironakaExpandF}
\end{equation}
On a given algebraic sheet of the generic fiber, the secondaries take definite values, so the restriction of $F$ to a sheet becomes an ordinary function of the primaries,
\begin{equation}
F\longrightarrow F_\beta(p),
\qquad \beta=1,\dots,N_S,
\label{eq:FtoFbeta}
\end{equation}
at least after passing to the corresponding branch of the cover. Accordingly, after changing variables from the original matrix entries to invariant variables, the matrix integral reorganizes, on the generic locus, into a finite sum of branchwise contributions of the form
\begin{equation}
Z\sim \sum_\beta \int \prod_{A=1}^{N_P}\dd p_A\,I_\beta(p_A).
\label{eq:branchwiseGeneralAgain}
\end{equation}
In the real Hermitian problem one must further impose positivity conditions, and these determine which algebraic branches are directly realized on the original contour\footnote{Saddles that only contribute after the original problem is complexified can still play an important role, as is the case for instantons.}. Nevertheless, the basic structure is universal: the primaries provide the continuous variables, while the secondaries encode the discrete fiber data.

From the physical point of view this distinction is important. A single perturbative expansion explores only one local branch of the invariant geometry. The full finite algebraic cover is therefore invisible in a strictly perturbative treatment. The secondaries are precisely the invariants that remember this global discrete structure. This is the first indication that they should be associated not with additional perturbative oscillators, but with sector data of a genuinely non-perturbative kind.

\section{Branch contributions as saddles in invariant space}
\label{sec:saddles}

Our ultimate goal is to develop evidence that the primary invariants describe the perturbative degrees of freedom of the matrix model, while the secondary invariants are associated with non-perturbative degrees of freedom. The finite algebraic cover identified in the previous section already suggests such an interpretation: the primaries parametrize the continuous base, while the secondaries resolve the discrete fiber lying above it. It is important, however, to distinguish the exact algebraic organization of the quotient from the dynamical question of how these sectors enter the path integral. The branch decomposition identified above was obtained from the exact change of variables to invariant coordinates. The question we address in this section is whether one can formulate an invariant-space variational problem whose critical locus reproduces this algebraic branch structure.

For two matrices, the quotient has only a single branch. After changing variables to invariants, the integral reduces to an ordinary integral over the primary variables alone. In this case, the primary invariants are not only the natural coordinates on the quotient, but also the variables that are actually integrated over in the path integral, so that they indeed play the role of perturbative degrees of freedom.

For three matrices the new ingredient is the oriented volume $\tau$ of \eqref{eq:tauDef}, or equivalently the imaginary part of the non-trivial secondary $s_1$, as in \eqref{eq:ImTraceTau}. The branch structure is controlled by the relation
\begin{equation}
\tau^2=\det G,
\label{eq:tauAgain}
\end{equation}
with $G$ the Gram matrix built from the primary invariants. Thus, once the primaries are fixed, there are two branches, corresponding to the two possible signs of $\tau$. Since the measure \eqref{eq:threeMeasureWithSO3} depends on $\tau=\sqrt{\det G}$, the measure for the two branches is determined by these signs. This two-branch structure can be obtained from a simple effective action in invariant variables. Introducing a Lagrange multiplier $\lambda$ and an extra variable $\tau$, one may write
\begin{equation}
S_{\rm eff}^{(3)}(p,\tau,\lambda)=S_{\rm mod}^{(3)}(p)-i\lambda\bigl(\tau^2-\det G(p)\bigr),\label{eq:Seff3}
\end{equation}
where $S_{\rm mod}^{(3)}(p)$ denotes the reduced action along the primary directions. With this effective action the matrix integral becomes
\begin{equation}
Z=\frac{\pi^2}{32}\int_{G\succeq 0} \dd^9 p \int_{-\infty}^\infty \frac{\dd\lambda}{2\pi}\int_{-\infty}^\infty \dd\tau \,\,e^{-S_{\rm eff}^{(3)}(p,\tau,\lambda)}
\end{equation}
Performing the integral over $\lambda$ produces a delta function. Using the delta function to perform the integral over $\tau$ produces the $1/\sqrt{\det G}$ dependence in the measure as well as the sum over branches, demonstrating the equivalence to our original matrix integral.

We will now perform the integral over the auxilliary $\lambda,\tau$ variables using the saddle point approximation. The saddle equation for $\lambda$ imposes
\begin{equation}
\tau^2=\det G(p),
\label{eq:tauConstraintFromLambda}
\end{equation}
while the saddle equation for $\tau$ gives
\begin{equation}
-2i\lambda\tau=0.
\label{eq:tauSaddleEqn}
\end{equation}
On the generic locus where $\tau\neq 0$, one obtains $\lambda=0$. The critical set therefore splits into two components,
\begin{equation}
\tau=\pm\sqrt{\det G(p)},\label{eq:twoCriticalComponents}
\end{equation}
which are precisely the two branches found in the explicit change of variables. In this sense the two branch contributions are genuine saddles of the effective invariant-space action.

The four-matrix example is richer, and this is the first example for which the distinction between algebraic sheets and the real Hermitian contour appears. The even quotient is determined by the quartic equation $\Phi(p,\Delta_{(4)})=0$,
while the cubic invariants satisfy further algebraic relations with the primaries and with $\Delta_{(4)}$. Over the complexified quotient, these relations define a finite algebraic cover of generic degree eight. In the real Hermitian problem one must then intersect this cover with the positivity domain of the Gram matrix so the full algebraic sector structure is generically larger than the set of sheets directly visible on the original real contour.

Let $T=(T_{123},T_{124},T_{134},T_{234})$, and define the four polynomial constraints
\begin{align}
F_1(p,\Delta_{(4)},T)&\equiv T_{123}^2-C_{44}(p,\Delta_{(4)})=0,\label{eq:F1poly}\\
F_2(p,\Delta_{(4)},T)&\equiv T_{123}T_{124}-\sigma_{124}\,C_{34}(p,\Delta_{(4)})=0,\label{eq:F2poly}\\
F_3(p,\Delta_{(4)},T)&\equiv T_{123}T_{134}-\sigma_{134}\,C_{24}(p,\Delta_{(4)})=0,\label{eq:F3poly}\\
F_4(p,\Delta_{(4)},T)&\equiv T_{123}T_{234}-\sigma_{234}\,C_{14}(p,\Delta_{(4)})=0,\label{eq:F4poly}
\end{align}
where the signs $\sigma_{124},\sigma_{134},\sigma_{234}=\pm1$ are fixed by the conventions used in the branch formulas of section~\ref{sec:four}. For fixed $(p,\Delta_{(4)})$ these equations have, on the generic locus, exactly two solutions, corresponding to the two choices of sign of $T_{123}$. Together with the four roots of $\Phi(p,\Delta_{(4)})=0$, they describe all eight branches at once. The exact matrix integral can be written as
\begin{eqnarray}
Z&=&\frac{\pi^2}{4}\int d^{13}p\,\int d\Delta_{(4)}\,\int d^4T\; \Theta\bigl(G(p,\Delta_{(4)})\succeq0\bigr)\,
\delta\bigl(\Phi(p,\Delta_{(4)})\bigr)\,\mathcal J_T(p,\Delta_{(4)},T)\cr\cr
&&\qquad\times\prod_{a=1}^4 \delta\bigl(F_a(p,\Delta_{(4)},T)\bigr)\,
e^{-S_{\rm mod}^{(4)}(p,\Delta_{(4)},T)},\label{eq:fourExactAll8}
\end{eqnarray}
where
\begin{equation}
\mathcal J_T(p,\Delta_{(4)},T)=\left|\det\frac{\partial(F_1,F_2,F_3,F_4)}{\partial(T_{123},T_{124},T_{134},T_{234})}
\right|\label{eq:JTdef}
\end{equation}
is the Jacobian for the change from the cubic variables to the polynomial constraints. For the choice \eqref{eq:F1poly}--\eqref{eq:F4poly}, one finds $\mathcal J_T=2\,|T_{123}|^4$ on the generic locus. The representation \eqref{eq:fourExactAll8} is exact. Indeed, for fixed $(p,\Delta_{(4)})$ the identity
\begin{equation}
\mathcal J_T\,\prod_{a=1}^4\delta\bigl(F_a(p,\Delta_{(4)},T)\bigr)=\sum_{\varepsilon=\pm1}
\delta^{(4)}\!\Bigl(T-T^{(\varepsilon)}(p,\Delta_{(4)})\Bigr)\label{eq:deltaTresolve}
\end{equation}
localizes the $T$-integral on the two cubic branches lying above $(p,\Delta_{(4)})$. The remaining $\Delta_{(4)}$ integral is then localized by
\begin{equation}
\delta\bigl(\Phi(p,\Delta_{(4)})\bigr)=\sum_{r=1}^4\frac{\delta\bigl(\Delta_{(4)}-\Delta_{(4),r}(p)\bigr)}
{\left|\partial_{\Delta_{(4)}}\Phi\bigl(p,\Delta_{(4),r}(p)\bigr)\right|},
\label{eq:deltaPhiResolveAgain}
\end{equation}
so that one recovers the sum over the eight algebraic branches with the correct measure,
\begin{equation}
Z=\frac{\pi^2}{4}\int d^{13}p\sum_{r=1}^4\sum_{\varepsilon=\pm1}\frac{\Theta_r(p)}
{\left|\partial_{\Delta_{(4)}}\Phi\bigl(p,\Delta_{(4),r}(p)\bigr)\right|}\exp\Bigl[-S_{{\rm mod},r,\varepsilon}^{(4)}(p)\Bigr].
\label{eq:recover8branches}
\end{equation}
To exhibit the same structure as a saddle problem, write the delta functions in Fourier form
\begin{equation}
Z=\frac{\pi^2}{4}\int d^{13}p\,\int d\Delta_{(4)}\,\int d^4T\,\int\frac{d\lambda_\Phi}{2\pi}\,\prod_{a=1}^4\int\frac{d\mu_a}{2\pi}\;\Theta\bigl(G(p,\Delta_{(4)})\succeq0\bigr)\,e^{-S_{\rm eff}^{(4)}},\label{eq:fourAll8Fourier}
\end{equation}
with
\begin{equation}
S_{\rm eff}^{(4)}=S_{\rm mod}^{(4)}(p,\Delta_{(4)},T)-\log \mathcal J_T(p,\Delta_{(4)},T)-i\lambda_\Phi\,\Phi(p,\Delta_{(4)})-i\sum_{a=1}^4 \mu_a\,F_a(p,\Delta_{(4)},T).\label{eq:Seff4All8}
\end{equation}
We will do the integrals over $\lambda_\Phi$, $T$, $\mu_a$ and $\Delta_{(4)}$ using a saddle point analysis. The saddle-point equations for $\lambda_\Phi$ and $\mu_a$ give
\begin{equation}
\Phi(p,\Delta_{(4)})=0,\qquad F_a(p,\Delta_{(4)},T)=0,\qquad a=1,\dots,4.\label{eq:multiplierEqsAll8}
\end{equation}
These are exactly the algebraic branch equations, so every generic branch point lifts to a saddle of the enlarged auxiliary system. For fixed $p$, consider a point $(\Delta_{(4)},T)$ on one of the eight branches, so that \eqref{eq:multiplierEqsAll8} holds. On the generic locus,
\begin{equation}
\partial_{\Delta_{(4)}} \Phi(p,\Delta_{(4)})\neq 0,\qquad\det\frac{\partial(F_1,F_2,F_3,F_4)}
{\partial(T_{123},T_{124},T_{134},T_{234})}\neq 0.\label{eq:genericLocusInvertible}
\end{equation}
The stationarity equations with respect to $\Delta_{(4)}$ and the four cubic variables then read
\begin{align}
0&=\partial_{\Delta_{(4)}} S_{\rm eff}^{(4)}=\partial_{\Delta_{(4)}}\Bigl(S_{\rm mod}^{(4)}-\log\mathcal J_T\Bigr)
-i\lambda_\Phi\,\partial_{\Delta_{(4)}}\Phi-i\sum_{a=1}^4 \mu_a\,\partial_{\Delta_{(4)}} F_a,
\label{eq:DeltaStationaryAll8}\\
0&=\partial_{T_I} S_{\rm eff}^{(4)}=\partial_{T_I}\!\Bigl(S_{\rm mod}^{(4)}-\log\mathcal J_T\Bigr)
-i\sum_{a=1}^4 \mu_a\,\partial_{T_I}F_a,\qquad I\in\{123,124,134,234\}.\label{eq:TStationaryAll8}
\end{align}
The Jacobian matrix $\partial_{T_I}F_a$ is invertible on the generic locus, so that the equations \eqref{eq:TStationaryAll8} determine the multipliers $\mu_a$ uniquely. Once these are fixed, \eqref{eq:DeltaStationaryAll8} determines $\lambda_\Phi$ uniquely because $\partial_{\Delta_{(4)}}\Phi\neq0$. In this precise sense, all eight algebraic branches arise as saddles of a single invariant-space variational problem. The real Hermitian contour is obtained only after imposing the positivity condition $G(p,\Delta_{(4)})\succeq0$. Thus the auxiliary saddle problem naturally sees the full complexified eight-sheeted cover, while the original real contour selects the subset of saddles that lie on the physical Hermitian slice.

Finally, we return to the model with $S_N$ symmetry. The integral written in invariant variables \eqref{SNmodelInvInt} decomposes into a sum over $N!$ branches. We now show that this sum can be recast as a single integral governed by an effective action, with the $N!$ branches emerging as saddle points of that action. Following the strategy used above, we introduce additional auxiliary variables. The first is a Lagrange multiplier $\lambda$. The second is a variable constructed from the secondary invariants. Concretely, we choose a mixed invariant $s$ with the property that, for generic values of the primary invariants, its value is different on each of the $N!$ sheets. For example, one may take a generic linear combination of mixed invariants
\begin{equation}
s=\sum_{a,b} c_{ab}\,s_{ab},\qquad s_{ab}=\sum_{i=1}^N x_i^a y_i^b.
\end{equation}
For generic coefficients $c_{ab}$, this separates the sheets on the generic locus. Then, over a fixed point in the primary space, the $N!$ branch values are $s=s^{(\sigma)}(p)$ with $\sigma\in S_N$. Define the monic degree-$N!$ polynomial
\begin{equation}
P(p,s)=\prod_{\sigma\in S_N}\Bigl(s-s^{(\sigma)}(p)\Bigr).
\end{equation}
This polynomial has coefficients symmetric separately in the $x$-roots and in the $y$-roots, hence its coefficients are functions of the primaries $p$. Therefore $P(p,s)=0$ is an algebraic equation in invariant variables only. Its roots are exactly the sheets.

For a fixed point in primary space, we have the elementary identity
\begin{equation}
\delta\bigl(P(p,s)\bigr)\,\bigl|\partial_s P(p,s)\bigr|=\sum_{\sigma\in S_N}\delta\bigl(s-s^{(\sigma)}(p)\bigr),
\end{equation}
provided the roots are simple, which is true on the generic locus. Therefore
\begin{equation}
\sum_{\sigma\in S_N} F\bigl(p,s^{(\sigma)}(p)\bigr)=\int_{\mathbb R} ds\;\delta\bigl(P(p,s)\bigr)\,\bigl|\partial_s P(p,s)\bigr|\,F(p,s).
\end{equation}
Applying this to our Gaussian integral, we find
\begin{equation}
Z=\int_{\Gamma} d^{2N}p \int_{\mathbb R} ds\;\frac{\bigl|\partial_s P(p,s)\bigr|}{N!\,\sqrt{\Delta_x(p)\,\Delta_y(p)}}\;
\delta\bigl(P(p,s)\bigr)\;\exp\left[-\frac12\bigl(p_2^{(x)}+p_2^{(y)}\bigr)\right].
\end{equation}
This is a single integral over invariant variables, with the branch structure encoded in the algebraic constraint $P(p,s)=0$. Finally, we may write
\begin{equation}
Z=\int_{\Gamma} d^{2N}p \int_{\mathbb R} ds \int_{\mathbb R}\frac{d\lambda}{2\pi}\,\,\frac{\bigl|\partial_s P(p,s)\bigr|}{N!\,\sqrt{\Delta_x(p)\,\Delta_y(p)}}\;\;e^{-S_{\rm eff}(p,s,\lambda)},
\end{equation}
with effective action
\begin{eqnarray}
S_{\rm eff}(p,s,\lambda)&=&\frac12\bigl(p_2^{(x)}+p_2^{(y)}\bigr)-i\lambda P(p,s).
\end{eqnarray}

We now want to understand why the sheets become saddle points. We will do the integral over $s$ and $\lambda$ for some fixed values of the primaries. The effective action is
\begin{equation}
S_{\rm eff}(p,s,\lambda)= - i\lambda P(p,s)+\cdots,
\end{equation}
up to slowly varying measure terms and terms independent of $s$ and $\lambda$. Its stationary equations are
\begin{equation}
\frac{\partial S_{\rm eff}(p,s,\lambda)}{\partial \lambda}=-i\,P(p,s)=0\qquad
\frac{\partial S_{\rm eff}(p,s,\lambda)}{\partial s}=-i\,\lambda\,\partial_s P(p,s)=0.
\end{equation}
On the generic locus, the roots are simple, so $\partial_s P\bigl(p,s^{(\sigma)}(p)\bigr)\neq 0$ and hence the second equation implies $\lambda=0$. The first equation gives $P(p,s)=0$. Therefore the saddle points are exactly
\begin{equation}
(s,\lambda)=\bigl(s^{(\sigma)}(p),0\bigr),\qquad\sigma\in S_N.
\end{equation}
Thus, for each generic primary point $p$, the $N!$ sheets are the $N!$ isolated saddles of the auxiliary invariant action.

This is the statement we need for the present paper. The secondaries do not introduce additional continuous directions. Rather, they encode the discrete algebraic sectors singled out by the invariant-space critical locus. In the real Hermitian problem the original contour selects a distinguished real cycle inside this larger complexified geometry, so not every algebraic sheet need contribute directly as an independent real saddle. But from the point of view of the invariant formulation, all of these sheets are natural candidate sectors, and it is entirely plausible that they become relevant through contour decomposition, analytic continuation, or complex saddle phenomena.

This is precisely the structure one would like if the Hironaka decomposition is to play a physical role. In the oscillator language the secondaries generate distinguished seed states, while the primaries build towers of excitations above them. In the invariant integral the same pattern reappears in a different guise: the primaries control the continuous variables, while the secondaries label the discrete algebraic sectors of the invariant geometry. This is the sense in which they naturally suggest non-perturbative degrees of freedom.

\section{Discussion}\label{sec:discussion}

The Hironaka decomposition points to a remarkably suggestive organization of the Hilbert space. Rather than a single perturbative vacuum, one is led to a finite collection of non-perturbative sectors, each supporting its own Fock-like tower of excitations. In this interpretation, the primary invariants act as the oscillators generating perturbative states, while the secondary invariants label the sector data on which these towers are built. The aim of this paper has been to show that this structure is already visible in simple matrix integrals after passing to invariant variables.

The basic lesson of the explicit examples we considered is clear. For the $N=2$ matrix models the change of variables from matrix entries to invariant coordinates reorganizes the quotient over primary space by a finite algebraic fiber. Exactly the same reorganization occurs when rewriting the coordinate space $(\mathbb R^2)^N$ of $N$ bosons in terms of permutation invariant variables. The primary invariants furnish the continuous coordinates on the base, while the secondary invariants encode the discrete data that separates the different sheets of the fiber. In the two-matrix example this structure is trivial. In the three-matrix example it becomes a genuine double cover distinguished by the orientation sign. In the four-matrix example the geometry is richer: the even quotient is controlled by a quartic equation for $\Delta$, and the odd cubic data resolves a further double cover. The natural object singled out by the invariant-theoretic analysis is therefore an eight-sheeted algebraic cover of the complexified primary space. Finally, for the $S_N$ invariant case we obtained an $N!$ sheeted algebraic cover of the primary space.

The description in terms of invariant variables is the correct setting in which to compare the quotient geometry to the Hironaka decomposition. The statement that the invariant ring is a free module of rank $N_S$ over the primary ring is an algebraic statement, and it is this algebraic statement that controls the generic degree of the complexified quotient over primary space. The real Hermitian problem is more restrictive because the original integration contour imposes positivity conditions on the Gram data. As a result, the number of real sheets visible directly on the original contour need not coincide pointwise with the algebraic degree of the cover. This does not weaken the physical relevance of the secondaries. On the contrary, it sharpens it. The secondaries are naturally associated not with the accidental topology of a chosen real slice, but with the full finite algebraic sector structure of the invariant quotient.

This distinction is precisely what makes the secondaries natural candidates for non-perturbative degrees of freedom. A perturbative expansion is local: it probes one chosen branch and the continuous fluctuations around it. The full finite algebraic fiber is therefore invisible in a strictly perturbative analysis. The secondaries are the invariants that remember this extra structure. They do not behave like additional perturbative oscillators. Rather, they encode the algebraic sectors that must be included in order to reconstruct the full quotient. In this sense they are naturally associated with non-perturbative sector data.

The invariant-space actions introduced in this paper support this interpretation. They show that the algebraic branch relations are not merely kinematical artifacts of a change of coordinates. They arise as critical-locus equations of a natural auxiliary variational problem in invariant space. This does not mean that every algebraic sheet contributes as an independent real saddle of the original Hermitian matrix integral. The correct picture is subtler: the Hermitian contour should be regarded as a distinguished real cycle inside the complexified invariant geometry. From this point of view, the additional algebraic sheets are still physically relevant, because non-perturbative effects are well known to be controlled by complex saddles\cite{Dorigoni:2014hea}, contour decompositions and Stokes phenomena, not only by real stationary points on the original contour. The picture that emerges is striking. For the $d$-matrix model, the ring of invariants contains $1+(d-1)N^2$ primary invariants, while the number of secondary invariants grows as $e^{cN^2}$, with $c=O(1)$. Thus the matrix-model integral is an integral over $O(N^2)$ invariant variables, together with an integrand whose complexified saddle structure is exponentially large, with $e^{cN^2}$ saddles. 

There are several natural directions in which the present analysis should be extended. It would be valuable to develop the finite-cover picture systematically for larger values of $N$ and for larger numbers of matrices. The $N=2$ examples studied here are tractable enough to make the geometry explicit, but the real interest of the Hironaka decomposition in physics is likely to emerge in problems where the number of secondaries grows rapidly. To study higher values of $N$ it may be useful to study vector models~\cite{deMelloKoch:2025cec,Sundborg:2026cut}, which are more tractable. Further, one would like to understand the branch locus itself. The generic description in terms of a finite algebraic cover necessarily breaks down on discriminant loci where sheets collide, where the fiber becomes non-reduced, or where a chosen local generator no longer separates the branches. These loci are mathematically natural and physically suggestive. They are the places where different sector descriptions meet, and they may encode transitions between different effective invariant-space descriptions.

The relation to genuine non-perturbative physics should be developed more directly. In the present paper we have identified the finite algebraic sector structure and exhibited an invariant-space action whose critical locus reproduces it. The next step is to understand how the Hermitian contour decomposes inside the complexified quotient, and whether the secondary sectors can be related to Lefschetz thimbles, resurgence data, or other standard signatures of non-perturbative physics.

It is both important and interesting to move beyond zero-dimensional matrix integrals and study matrix quantum mechanics or quantum field theories in which the same invariant-theoretic structures are present. In such settings one could ask whether the sector decomposition identified here survives in a genuinely dynamical context, and whether the discrete fiber data encoded by the secondaries can be related more directly to background geometries, black-hole microstates, or superselection data. In particular, non-perturbative finite-$N$ effects play a central role in the description of black hole microstates, through the fortuity mechanism~\cite{Chang:2022mjp,Choi:2022caq,Choi:2023znd,Chang:2023zqk,Choi:2023vdm,Chang:2024zqi,Chen:2025sum,Kim:2025vup,Belin:2025hsg,Behan:2025hbx}: the microstates of $\frac{1}{16}$-BPS black holes fall into two classes, monotone and fortuitous. Monotone states remain BPS at all $N$, while fortuitous states lose their BPS nature above a critical $N$. It is important to understand how and if these non-perturbative effects can be incorporated in the framework proposed in this paper.

\begin{center} 
{\bf Acknowledgements}
\end{center}
We would like to thank Antal Jevicki for very useful discussions.
The work of RdMK is supported by a start up research fund of Huzhou University, a Zhejiang Province talent award and by a Changjiang Scholar award. The work of JPR is supported by the National Institute for Theoretical and Computational Sciences, NRF Grant Number 65212.

\begin{appendix}

\section{Comments on the bases of invariants}\label{invariantbases}

An essential input in our analysis is the system of primary and secondary invariants of the invariant ring. In this section we describe how the set of invariants employed in our computations is determined.

A basic quantity we can compute is the Hilbert series. The Hilbert series $H(x)$ is a rational function of $x$. When expanded as a power series, the coefficient of $x^n$ counts how many degree $n$ invariants there are. For an invariant ring that admits a Hironaka decomposition, the Hilbert series can always be manipulated into the form
\begin{equation}
H(x) = \frac{1 + \sum_i c^s_i x^i}{\prod_j (1 - x^j)^{c^m_j}}. \label{illpf}
\end{equation}
Every primary invariant is associated with a factor in the denominator and every secondary invariant is associated with a term in the numerator. The factor $(1 - x^j)$ in the denominator encodes a primary invariant of degree $j$. A term $x^j$ in the numerator encodes a secondary invariant of degree $j$. Thus, once the Hilbert series is computed, one has a count of the primary and secondary invariants as well as their degrees. One complication that comes up in practise is that one compute the rational function $H(x)$ directly, not its numerator and denominator. There is some ambiguity in passing from $H(x)$ to a numerator and a denominator simply because common factors may have been cancelled. In addition, the choice of the primary and secondary invariants is not unique.

For the two matrix model the Hilbert series is
\begin{equation}
H(x) = \frac{1}{(1-x)^2(1-x^2)^3}  \label{exactN2M2}
\end{equation}
indicating that there is a single degree zero secondary invariant and two degree 1 and three degree 2 primary invariants. The invariants in \eqref{eq:twoPrimariesList} are consistent with this count. This system of invariants was derived in~\cite{FHL,deMelloKoch:2025ngs}.

For the three matrix model the Hilbert series is
\begin{equation}
H(x) = \frac{1 + x^3}{(1-x)^3(1-x^2)^6}\label{exactN2M3}
\end{equation}
so there are two secondary invariants and nine primary invariants. This is again consistent with our choice \eqref{eq:threePrimariesList} and \eqref{twomatsecondaries}. This system of invariants is derived in~\cite{Teranishi,deMelloKoch:2025ngs}.

The Hilbert series for the four matrix model is given by
\begin{equation}
H(x) = \frac{1 + x^2 + 4x^3 + x^4 + x^6}{(1 - x)^4 (1 - x^2)^9},
\end{equation}
We see that there are four primary invariants of degree one and nine primary invariants of degree two, which agrees with our choice \eqref{eq:fourPrimariesExplicit}. In addition, there is a single secondary invariant of degree zero, two, four and six, and four secondary invariants of degree three, which agrees with our choice \eqref{eq:fourSecondaries}. This choice of invariants does not, as far as we are aware, appear in the literature. It is therefore necessary to confirm that this is a valid set of invariants.

The fact that our generators reproduce the degree structure predicted by the Hilbert series ensures that they produce the correct number of invariants at each degree. The remaining issue is to verify that these invariants are indeed free, so that no unexpected relations remain among them. Fortunately, this can be checked efficiently using the numerical method described in~\cite{deMelloKoch:2025eqt}. Since any possible relation must involve invariants of the same total degree, one proceeds degree by degree. For a fixed degree, we first construct the complete set of invariants generated by our chosen primaries and secondaries. In accordance with the Hironaka decomposition, each such invariant contains exactly one secondary, multiplied by an appropriate polynomial in the primaries. We then form the most general linear combination of these invariants, with undetermined coefficients, and evaluate it on randomly chosen numerical values of the underlying matrices. Repeating this procedure for sufficiently many independent choices of matrices produces a linear system for the coefficients. Once the number of independent equations matches the number of unknowns, we test for the existence of a non-trivial solution. The absence of such a solution implies that no relation is present, and hence that the proposed set of invariants is freely generated.

For the set of invariants given in \eqref{eq:fourPrimariesExplicit} and \eqref{eq:fourSecondaries}, we have checked explicitly that no non-trivial relations appear up to degree 12. This provides strong evidence that the proposed system of generators is correct.

\section{An Integral Identity}\label{IntegralDone}

In this Appendix we want to derive the identity
\begin{equation}
\int_{A\succ 0} d^6A\,\frac{e^{-\Tr A}}{\sqrt{\det A}}=\frac{\pi^{5/2}}{2}.
\end{equation}
Here $A$ is a real symmetric positive-definite $3\times 3$ matrix,
\begin{equation}
A=
\begin{pmatrix}
A_{11} & A_{12} & A_{13}\\
A_{12} & A_{22} & A_{23}\\
A_{13} & A_{23} & A_{33}
\end{pmatrix},
\qquad d^6A=dA_{11}\,dA_{12}\,dA_{13}\,dA_{22}\,dA_{23}\,dA_{33}.
\end{equation}
A convenient parametrization of the positive-definite symmetric matrix $A$ is
\begin{equation}
A=
\begin{pmatrix}
r_1^2 & r_1 a & r_1 c\\[3pt]
r_1 a & a^2+b^2 & ac+bd\\[3pt]
r_1 c & ac+bd & c^2+d^2+e^2
\end{pmatrix},
\qquad
r_1>0,\quad b>0,\quad e>0,\quad a,c,d\in\mathbb R.\label{eq:paramA}
\end{equation}
This is the Cholesky-type parametrization appropriate to the adapted coordinates we used in our examples. Every $A\succ 0$ can be written uniquely in this form. One recovers the parameters from $A$ as follows:
\begin{align}
r_1&=\sqrt{A_{11}},\qquad a=\frac{A_{12}}{r_1},\qquad c=\frac{A_{13}}{r_1},\\[3pt]
b&=\sqrt{A_{22}-a^2},\qquad d=\frac{A_{23}-ac}{b},\\[3pt]
e&=\sqrt{A_{33}-c^2-d^2}.
\end{align}
Because $A\succ 0$, all principal minors are positive, so $r_1>0$, $b>0$, and $e>0$. The Jacobian of this change of coordinates is
\begin{equation}
\left|\frac{\partial(A_{11},A_{12},A_{13},A_{22},A_{23},A_{33})}{\partial(r_1,a,b,c,d,e)}\right|=8r_1^3b^2e,
\end{equation}
so that
\begin{equation}
d^6A=8r_1^3b^2e\;dr_1\,da\,db\,dc\,dd\,de.\label{eq:d6A}
\end{equation}
The matrix \eqref{eq:paramA} is the Gram matrix of the three vectors
\begin{equation}
\vec m_1=(r_1,0,0),\qquad\vec m_2=(a,b,0),\qquad\vec m_3=(c,d,e),
\end{equation}
so its determinant is
\begin{equation}
\det A = \bigl(\det(\vec m_1,\vec m_2,\vec m_3)\bigr)^2= (r_1be)^2= r_1^2b^2e^2\label{eq:detA}
\end{equation}
and its trace is
\begin{equation}
\Tr A = r_1^2 + (a^2+b^2) + (c^2+d^2+e^2)= r_1^2+a^2+b^2+c^2+d^2+e^2.\label{eq:traceA}
\end{equation}
Using \eqref{eq:d6A} and \eqref{eq:detA},
\begin{equation}
\frac{d^6A}{\sqrt{\det A}}=\frac{8r_1^3b^2e}{r_1be}\,dr_1\,da\,db\,dc\,dd\,de=8r_1^2b\;
dr_1\,da\,db\,dc\,dd\,de.\label{eq:measureReduced}
\end{equation}
Substituting \eqref{eq:traceA} and \eqref{eq:measureReduced} into the integral gives
\begin{align}
\int_{A\succ 0} d^6A\,\frac{e^{-\Tr A}}{\sqrt{\det A}}&=8\int_0^\infty dr_1\, r_1^2 e^{-r_1^2}
\int_{-\infty}^{\infty} da\, e^{-a^2}\int_0^\infty db\, b\, e^{-b^2}\nonumber\\
&\qquad\times\int_{-\infty}^{\infty} dc\, e^{-c^2}\int_{-\infty}^{\infty} dd\, e^{-d^2}\int_0^\infty de\, e^{-e^2}.
\label{eq:factorized}
\end{align}
Each factor is elementary and the identity follows.

\end{appendix}

\end{document}